\documentclass[%
 aip,
 amsmath,amssymb,
 preprint,%
]{revtex4-1}

\usepackage{graphicx}
\usepackage{dcolumn}
\usepackage{bm}

\usepackage[utf8]{inputenc}
\usepackage[T1]{fontenc}
\usepackage{mathptmx}
\usepackage{etoolbox}

\usepackage{amsmath}
\usepackage{amsfonts}
\DeclareMathAlphabet\mathbfcal{OMS}{cmsy}{b}{n}

\usepackage{siunitx}
\usepackage{enumitem}
\usepackage[colorlinks=true, allcolors=blue]{hyperref}
\usepackage{multirow}
\usepackage{array}

\newcommand*\chem[1]{\ensuremath{\mathrm{#1}}}

\usepackage{soul} 

\newcommand{\commentout}[1]{}

\makeatletter
\def\@email#1#2{%
 \endgroup
 \patchcmd{\titleblock@produce}
  {\frontmatter@RRAPformat}
  {\frontmatter@RRAPformat{\produce@RRAP{*#1\href{mailto:#2}{#2}}}\frontmatter@RRAPformat}
  {}{}
}%
\makeatother
\begin{document}

\preprint{AIP/123-QED}

\title[]{Thermodynamically consistent incorporation of the Langmuir adsorption model into compressible fluctuating hydrodynamics}

\author{Hyun Tae Jung}
\affiliation{Department of Chemistry, Korea Advanced Institute of Science and Technology, Daejeon 34141, South Korea}

\author{Hyungjun Kim}
\affiliation{Department of Chemistry, Korea Advanced Institute of Science and Technology, Daejeon 34141, South Korea}

\author{Alejandro L. Garcia}
\affiliation{Department of Physics and Astronomy, San Jose State University, San Jose, California 95192, USA}

\author{Andrew J. Nonaka}
\affiliation{Center for Computational Sciences and Engineering, Lawrence Berkeley National Laboratory, Berkeley, California 94720, USA}

\author{John B. Bell}
\affiliation{Center for Computational Sciences and Engineering, Lawrence Berkeley National Laboratory, Berkeley, California 94720, USA}

\author{Ishan Srivastava}
\affiliation{Center for Computational Sciences and Engineering, Lawrence Berkeley National Laboratory, Berkeley, California 94720, USA}

\author{Changho Kim}
\email{ckim103@ucmerced.edu}
\affiliation{Department of Applied Mathematics, University of California, Merced, California 95343, USA}

\date{\today}

\begin{abstract}
For a gas-solid interfacial system where chemical species undergo reversible adsorption, we develop a mesoscopic stochastic modeling method that simulates both gas-phase hydrodynamics and surface coverage dynamics by coupling the Langmuir adsorption model with compressible fluctuating hydrodynamics.
To this end, we derive a thermodynamically consistent mass--energy update scheme that accounts for how the mass and energy variables in the gas and surface subsystems should be updated according to the changes in the number of molecules of each species in each subsystem due to adsorption and desorption events.
By performing a stochastic analysis for the ideal Langmuir model and the full hydrodynamic system, we analytically confirm that our mass--energy update scheme captures thermodynamic equilibrium predicted by equilibrium statistical mechanics.
We find that an internal energy correction term is needed, which is attributed to the difference in the mean kinetic energy of gas molecules colliding with the surface from that computed from the Maxwell--Boltzmann distribution.
By performing an equilibrium simulation study for an ideal gas mixture of \chem{CO} and \chem{Ar} with \chem{CO} undergoing reversible adsorption, we validate our overall simulation method and implementation.
\end{abstract}

\maketitle

\section{\label{sec:Introduction}Introduction}

Computational modeling of reactive gas-solid interfacial systems, e.g., heterogeneous catalysts~\cite{Deutschmann2011}, plays an important role in various fields of science and engineering, including energy and environmental sciences~\cite{JoshiNandakumar2015,GrajciarHeardBondarenkoPolynskiMeeprasertPidkoNachtigall2018}.
Due to its intrinsic multi-phase, multi-scale nature, an accurate and computationally efficient description of both the reactive dynamics on the catalytic surface and the transport dynamics in the gas phase is required.
However, since these dynamics have disparate natures, the use of a single traditional simulation approach usually leads to inaccurate or computationally inefficient simulations.
For example, while computational fluid dynamics (CFD)~\cite{MaestriCuoci2013, HettelWornerDeutschmann2018} provides a computationally efficient method to describe hydrodynamic behavior of the fluid phase, a simplified continuum representation of the catalytic surface may give inaccurate results because reaction kinetics based on the mean-field approximation (whether it is an empirical kinetic model or microkinetic model) may fail to provide an accurate description of surface catalytic reactions~\cite{SalciccioliStamatakisCaratzoulasVlachos2011, AndersenPlaisanceReuter2017, PratsIllasSayos2018}.
On the other hand, although particle-based simulation methods, such as molecular dynamics (MD)~\cite{MuellerVanDuinGoddard2010, SenftleHongIslamKylasaZhengShinJunkermeierEngelHervertJanikAktulgaVerstraelenGramavanDuin2016} and kinetic Monte Carlo (KMC)~\cite{AndersenPanosettiReuter2019, PinedaStamatakis2022}, can accurately model reactive dynamics on the surface, it is computationally prohibitive to simulate the entire interfacial system using them.

As an alternative, several hybrid simulation approaches have been proposed for reactive gas-solid interfacial systems.
In the CFD--KMC hybrid approach, CFD is employed for gas-phase hydrodynamics, whereas KMC is used for surface chemistry.
Most existing CFD--KMC hybrid simulation methods, see for example Refs.~\citenum{MajumderBroadbelt2006, MateraReuter2010, MeiLin2011, SchaeferJansen2013, SuttonLorenziKrogelXiongPannalaMateraSavara2018,YunTomOrkoulasChristofides2022}, are based on the macro-micro coupling structure or the heterogeneous multiscale method~\cite{EEngquistLiRenVandenEijnden2004, WeinanE2011}, where the KMC (micro solver) is \textit{passively} coupled to CFD (macro solver) using one-way coupling under the assumption of complete scale separation.
In other words, when the CFD solver needs surface reaction kinetics information at each point on the surface to update the (macroscopic) state of the system, the KMC solver is called to estimate that information.
However, for a mesoscale gas-solid interfacial system, where the gas-phase hydrodynamics and surface reaction dynamics have comparable time and length scales, this CFD--KMC hybrid approach based on the macro-micro coupling is not applicable and a new hybrid simulation approach based on \textit{two-way, concurrent} coupling is needed.

In this mesoscale hybrid approach, it is assumed that the domain of the surface chemistry solver corresponds to a physical boundary of the continuum hydrodynamics solver and these solvers update the states of the corresponding subsystems concurrently while exchanging molecules due to adsorption and desorption.
Since thermal fluctuations are significant at mesoscales, they need to be incorporated in the continuum CFD solver.
The fluctuating hydrodynamics (FHD) approach~\cite{LandauLifschitz1987, ZarateSengers2006, CroccoloZarateSengers2016} provides a suitable mesoscopic simulation framework~\cite{GarciaBellNonakaSrivastavaLadigesKim2024, BalakrishnanGarciaDonevBell2014, SrivastavaLadigesNonakaGarciaBell2023} and has been used to develop two-way, concurrent continuum--particle coupling methods for nonreactive hydrodynamic systems, e.g., coupling with MD~\cite{GiupponiDeFabritiisCoveney2007} or direct simulation Monte Carlo~\cite{DonevBellGarciaAlder2010}.
This paper aims to serve as a precursor for the development of an FHD--KMC coupling, which we believe is a promising hybrid simulation approach for reactive gas-solid interfacial systems at mesoscales.

The main theoretical challenge that this paper addresses is the development of a thermodynamically consistent continuum--particle coupling.
In the context of mesoscopic modeling, the correct description of thermal fluctuations throughout the overall system is of critical importance.
As mentioned above, the gas and solid subsystems exchange molecules via adsorption and desorption, the occurrences of which are modeled stochastically.
Hence, the states of the subsystems should be updated based on the stochastic quantities corresponding to the numbers of occurrences of adsorption and desorption. 
Assuming that thermal fluctuations are correctly described within each subsystem by the given continuum and particle-based descriptions (i.e., FHD and KMC), we derive an update scheme that guarantees the correct description of thermal fluctuations across the gas-solid interface.
To focus on the essential picture of the update scheme, we assume in this paper that there are no surface chemical reactions other than reversible adsorption reactions, and they follow the Langmuir model, undergoing molecular (i.e., one-site) adsorption~\cite{SwensonStadie2019}.
In this case, the mean-field description of the surface coverage dynamics becomes valid and equivalent to the KMC description.
This feature enables us to focus on theoretical development and perform an extensive computational validation study of our thermodynamically consistent update scheme without implementing the full FHD--KMC coupling.
In other words, while our update scheme is constructed for the FHD--KMC coupling, we first implement it on our existing FHD simulation codes~\cite{SrivastavaLadigesNonakaGarciaBell2023} using a mean-field description and test it thoroughly.
Implementation of the FHD--KMC coupling requires additional algorithmic components (e.g., efficient communication between the FHD and KMC solvers), which will be presented in a subsequent paper.
In addition, we believe that findings of this paper are applicable to other mesoscopic continuum--particle coupling approaches.

In this paper, to develop and validate our thermodynamically consistent continuum--particle coupling, we perform a systematic stochastic analysis both analytically and numerically. 
A similar approach to construct a thermodynamically consistent mesoscopic simulation methodology based on the continuum FHD description has been established for nonreactive fluid systems~\cite{DonevVanden-EijdenGarciaBell2010} and extended to reactive fluid systems~\cite{KimNonakaBellGarciaDonev2017, KimNonakaBellGarciaDonev2018, PolimenoKimBlanchetteSrivastavaGarciaNonakaBell2025}. 
This paper further extends this systematic approach to reactive interfacial systems by using the following strategies.
First, before analyzing the full dynamical system with gas-phase hydrodynamics, we consider the ideal Langmuir model in equilibrium, for which the state of the gas phase is well described by the instantaneous temperature and species mass densities, and construct a thermodynamically consistent mass--energy update.
Second, as a thermodynamic consistency criterion for constructing and validating our mass--energy update and overall continuum--particle coupling schemes, we use the fact that the resulting dynamical systems must reproduce the thermodynamic equilibrium predicted by equilibrium statistical mechanics.
In our analytical approach, we consider the weak-noise limit where the magnitude of instantaneous fluctuations in the state variables is relatively small.
In this limit, the time evolution equations can be reduced to the system of linear stochastic differential equations (SDEs) driven by additive Gaussian white noise, for which the analytic solutions are given as multivariate Gaussian processes.
To numerically confirm thermodynamic equilibrium, we compute the (co-)variances and static structure factor spectra of the state variables by performing equilibrium simulations.
Third, we develop a new thermodynamically consistent reaction (TCR) model for Langmuir adsorption.
For gas-phase reactions, the TCR model was introduced to ensure that the relationship between the equilibrium constant and the rate constants is preserved, which is crucial for thermodynamic consistency in reactive mesoscopic simulations~\cite{PolimenoKimBlanchetteSrivastavaGarciaNonakaBell2025}.
Thermodynamic equilibrium is determined by the chemical composition of the gas and the corresponding chemical potentials.
Hence, even if chemical potentials do not explicitly appear in the final form of a thermodynamically consistent numerical method, the formulation of the method and parameter selection should nevertheless be based on consistent chemical potential models.  
The TCR model approach guarantees that the resulting mass--energy update scheme is based on consistent chemical potential models for gas and adsorbate molecules.
The TCR model assumes a simpler form of the equilibrium constant and adsorption and desorption rate constants, based on the modified Arrhenius equation, which facilitates the analytical stochastic analysis.

The rest of the paper is organized as follows.
In Section~\ref{sec:MassEnergyUpdate}, we first consider the ideal Langmuir model in equilibrium and derive how instantaneous values of the mass and energy variables should be updated in terms of adsorption/desorption count.
In Section~\ref{sec:FullSystemDescription}, we then consider the full spatio-temporal evolution of the gas-solid interfacial system by incorporating the FHD description for the hydrodynamics of the gas subsystem.
In Section~\ref{sec:NumericalValidation}, we validate our theoretical formulation and numerical implementation by performing equilibrium simulations.
In Section~\ref{sec:Conclusion}, we conclude the paper with a summary and future work.

\section{\label{sec:MassEnergyUpdate}Mass--Energy Update}

\begin{figure}
\includegraphics[width=0.4\linewidth]{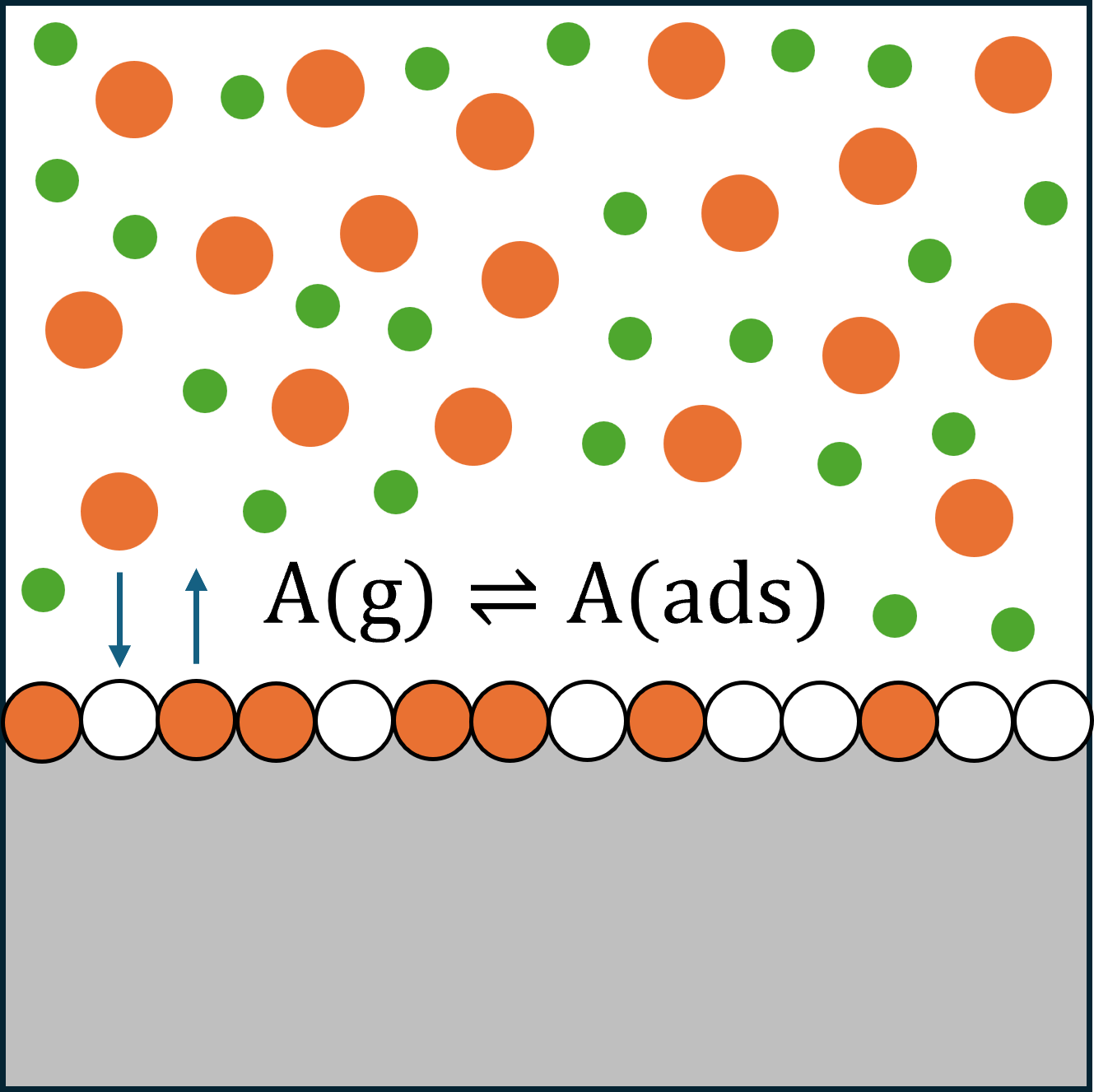}
    \caption{\label{Fig:LangmuirAdsoprtionModel}
    Ideal Langmuir model undergoing reversible molecular (i.e., one-site) adsorption.
    It consists of the gas subsystem containing an ideal gas mixture and the solid subsystem containing an ideal adsorbent. 
    We assume that the ideal gas mixture consists of a chemical species ($\chem{A}$, orange) undergoing reversible adsorption~\eqref{eq:revads} and a nonreactive species ($\chem{B}$, green).}
\end{figure}

In this section, we focus on the construction of a thermodynamically consistent update of the mass and energy variables.
To this end, we consider a gas-solid interfacial system in equilibrium, where the time evolution of its state can be described in terms of the mass and energy variables of the gas and solid subsystems.
More specifically, for the ideal Langmuir model~\cite{SwensonStadie2019, Hill1987}, we derive a mass--energy update scheme for the species mass densities (of chemical species undergoing reversible molecular adsorption), temperature of the gas phase, the corresponding surface coverages, and temperature of the solid phase due to adsorption and desorption events.
While our formulation can be readily extended to the multiple species case, for simplicity of the derivation, we assume that there is a single chemical species undergoing reversible adsorption and a single nonreactive species, see Figure~\ref{Fig:LangmuirAdsoprtionModel}.
Note that while we here derive the update scheme for a general case where the solid subsystem has finite heat capacity, we will consider the infinite heat capacity limit to assume constant surface temperature in Section~\ref{sec:FullSystemDescription}.

In Section~\ref{sec:TCRmodel}, we first construct a thermodynamically consistent reaction (TCR) model for Langmuir adsorption by investigating the temperature dependence of the equilibrium constant and rate constants for reversible adsorption.
In Section~\ref{sec:DerivationMassEnergyUpdate}, we derive a thermodynamically consistent mass--energy update for the ideal Langmuir model.
In Section~\ref{sec:PhysicalInterpretation}, we provide a physical interpretation of the energy correction term appearing in the mass--energy update.

\subsection{\label{sec:TCRmodel}TCR Model for Langmuir Adsorption}

As mentioned above, we consider the ideal Langmuir model undergoing reversible molecular (i.e., one-site) adsorption~\cite{SwensonStadie2019, Hill1987}:
\begin{equation}
\label{eq:revads}
    \chem{A}(g) + \varnothing \rightleftharpoons \chem{A}(ads), 
\end{equation}
where a gas molecule, $\chem{A}(g)$, is adsorbed onto an empty site $\varnothing$ on the surface or an adsorbed molecule, $\chem{A}(ads)$, is desorbed from the surface.
As shown in Figure~\ref{Fig:LangmuirAdsoprtionModel}, an ideal gas mixture of species $\chem{A}$ and $\chem{B}$ occupies the gas subsystem and there is a monolayer of adsorption sites on the surface of the solid subsystem.
The Langmuir model assumes that the occurrences of adsorption and desorption events on each site is independent of those on the other sites.
Since chemical equilibrium of species $\chem{A}$ is not affected by the presence of species $\chem{B}$, in Section~\ref{sec:MassEnergyUpdate}, we denote the mass density and partial pressure of species $\chem{A}$ by $\rho\equiv\rho_\chem{A}$ and $p\equiv p_\chem{A}$, respectively.
We assume that the system is in equilibrium at temperature $T$.

The kinetics of adsorption and desorption is described by surface coverage $\theta$, which is defined as the ratio of the number of occupied sites to that of total adsorption sites.
The mean rates of adsorption and desorption, $\bar{r}_a$ and $\bar{r}_d$, are given by
\begin{equation}
\label{eq:meanrates}
    \bar{r}_a = k_a p (1-\theta),\quad \bar{r}_d = k_d \theta,
\end{equation}
where $k_a$ and $k_d$ are the rate constants of adsorption and desorption.
Note that $k_a$ and $k_d$ are defined as constants (with respect to concentration variables, i.e., $p$ and $\theta$) in the rate expressions~\eqref{eq:meanrates} but they are assumed to be functions of temperature.
Since the equilibrium constant $K$ for \eqref{eq:revads} is defined as
\begin{equation}
    K = \frac{\theta}{p(1-\theta)}
    \label{eq:EquilibConstantDefn}
\end{equation}
and $\bar{r}_a = \bar{r}_d$ at equilibrium, it is easy to see
\begin{equation}
\label{eq:relKkakd}
    K(T) = \frac{k_a(T)}{k_d(T)}.
\end{equation}

To characterize the temperature dependence of the equilibrium constant and the adsorption and desorption rate constants~\cite{Kolasinski2012}, the TCR model approach~\cite{PolimenoKimBlanchetteSrivastavaGarciaNonakaBell2025} assumes that the temperature dependence of these constants is given in the form of the modified Arrhenius equation~\cite{Laidler1996, IUPAC1997}:
\begin{subequations}
\label{eq:modifiedArrhenius}
\begin{align}
    K(T) \sim T^{\beta_K} \exp\left(-\frac{\alpha_K}{k_B T}\right),\\
    k_a(T) \sim T^{\beta_a} \exp\left(-\frac{\alpha_a}{k_B T}\right),\\
    k_d(T) \sim T^{\beta_d} \exp\left(-\frac{\alpha_d}{k_B T}\right),
\end{align}
\end{subequations}
where $k_B$ is the Boltzmann constant.
Note that, if two of these constants are assumed to have this form, the other constant will also be in the form due to the relation~\eqref{eq:relKkakd}.
Furthermore, this relation implies that $(\alpha_a,\beta_a)$ and $(\alpha_d,\beta_d)$ cannot be chosen independently because they must satisfy
\begin{equation}
\label{eq:alphabetarel}
    \alpha_a - \alpha_d = \alpha_K, \quad \beta_a - \beta_d = \beta_K.
\end{equation}
Since the equilibrium constant is a thermodynamic quantity, the relations~\eqref{eq:alphabetarel} show that there is in fact only a single independent rate constant describing how the system relaxes to equilibrium.

In the TCR model approach, a modified Arrhenius form of $K(T)$ is obtained using a constant specific heat capacity assumption~\cite{PolimenoKimBlanchetteSrivastavaGarciaNonakaBell2025}.
We assume that the specific internal energies of an ideal gas molecule~$\chem{A}$ and an ideal adsorbate molecule~$\chem{A}$ are given as
\begin{align}
\label{eq:eg}
    &e_g(T) = \varepsilon_g + c_{v,g} T,\\
\label{eq:eads}
    &e_{ads}(T) = \varepsilon_{ads} + c_{ads}T,
\end{align}
where $\varepsilon_g$ is the specific internal energy of a gas molecule extrapolated to $T=0$, $c_{v,g}$ is the specific heat capacity of a gas molecule at constant volume, $\varepsilon_{ads}$ is the specific internal energy of an adsorbate molecule extrapolated to $T=0$, and $c_{ads}$ is the specific heat capacity of an adsorbate.
We can then derive the following expression of $K$, see Appendix~\ref{sec:appendixTCR} for details:
\begin{equation}
\label{eq:TCRKKst}
    K(T) = K(T^{st})
    \exp\left[\frac{m(\varepsilon_g-\varepsilon_{ads})}{k_B} \left(\frac{1}{T}-\frac{1}{T^{st}}\right)\right]\left(\frac{T}{T^{st}}\right)^{\frac{m}{k_B}(c_{ads}-c_{p,g})},
\end{equation}
where $m\equiv m_\chem{A}$ is the mass of a molecule~$\chem{A}$, $T^{st}$ is the standard (or reference) temperature, and $c_{p,g} \equiv c_{v,g} + k_B/m$ is the specific heat capacity of a gas molecule~$\chem{A}$ at constant pressure.
We thus obtain
\begin{equation}
\label{eq:alphaKbetaK}
    \alpha_K = m(\varepsilon_{ads} - \varepsilon_g) = -\Delta U, \quad
    \beta_K = \frac{m}{k_B}(c_{ads}-c_{p,g}).
\end{equation}
Here, we have introduced $\Delta U > 0$ as the surface binding energy of a molecule~$\chem{A}$~\cite{ReuterScheffler2006}.

\subsection{\label{sec:DerivationMassEnergyUpdate}Derivation of Mass--Energy Update}

\subsubsection{\label{sec:PhysicalModeling}Physical Modeling}

The state of the ideal Langmuir model, which is described by the mass density $\rho$ and temperature $T$ of the gas phase as well as the surface coverage $\theta$ and temperature $T_s$ of the surface phase, evolves in time due to adsorption and desorption events.
Since adsorption and desorption occurrences obey Poisson statistics, we introduce a notation $N_\lambda (t)$ to denote a Poisson process with a time-varying rate $\lambda$; 
for finite $\Delta t>0$, $\Delta N = N_\lambda(t+\Delta t)-N_\lambda(t)$ has a Poisson distribution with mean $\int_t^{t+\Delta t} \lambda d\tau$.
We denote Poisson processes for adsorption and desorption counts by $N_{\lambda_a}(t)$ and $N_{\lambda_d}(t)$, respectively.
From Eq.~\eqref{eq:meanrates}, the corresponding time-varying rates are given as
\begin{equation}
\label{eq:lambdaad}
    \lambda_a = k_a(T)p(\rho,T)(1-\theta)N_{tot}, \quad
    \lambda_d = k_d(T_s)\theta N_{tot},
\end{equation}
where $N_{tot}$ is the total number of adsorption sites and $p(\rho,T)$ is given by the ideal gas law:
\begin{equation}
    p(\rho,T) = \frac{k_B}{m}\rho T.
\end{equation}
Note that the instantaneous temperature $T$ of the gas is used for $k_a$, whereas the instantaneous temperature $T_s$ of the surface is used for $k_d$.

For a small finite time interval $\Delta t$, we denote the numbers of adsorption and desorption occurrences by $\Delta N_a$ and $\Delta N_d$, respectively.
Since state changes are proportional to the adsorption-desorption count defined as $\Delta N_{ad} = \Delta N_a - \Delta N_d$, we write the update scheme as
\begin{subequations}
\label{eq:massenergyupdate}    
\begin{align}
    &\rho(t+\Delta t) = \rho(t) - \frac{m}{V} \Delta N_{ad}, \\
    &T(t+\Delta t) = T(t) + \frac{\sigma q }{C} \Delta N_{ad}, \\
    &T_s(t+\Delta t) = T_s(t) + \frac{(1-\sigma)q}{C_s} \Delta N_{ad}, \\
    &\theta(t+\Delta t) = \theta(t) + \frac{1}{N_{tot}} \Delta N_{ad}.
\end{align}
\end{subequations}
Here, $V$ is the volume of the gas, $q$ is the heat of adsorption, $0<\sigma<1$ is the ratio indicating what fraction of the heat is absorbed by the gas phase during an adsorption event, and $C$ and $C_s$ are the heat capacities of the gas and solid subsystems.
Note that $q$ and $\sigma$ are not \textit{a priori} known and need to be determined from the adsorption and desorption rate parameters, $\alpha_a$, $\beta_a$, $\alpha_d$, and $\beta_d$. 
In the following, we perform a stochastic analysis to determine $q$ and $\sigma$ and confirm the thermodynamic consistency of the update scheme~\eqref{eq:massenergyupdate}.

\subsubsection{\label{sec:StochasticAnalysis}Stochastic Analysis}

As mentioned in the Introduction, to analytically investigate the mass-energy update~\eqref{eq:massenergyupdate}, we linearize it around the equilibrium state and consider the weak-noise limit to obtain a system of linear SDEs with additive Gaussian noise.
To this end, by assuming that the system is in equilibrium at temperature $\bar{T}$ and the equilibrium values of the gas mass density and surface coverage are $\bar{\rho}$ and $\bar{\theta}$, we rewrite the system in terms of the instantaneous fluctuations: $\delta \rho = \rho-\bar{\rho}$, $\delta T = T-\bar{T}$, $\delta T_s = T_s-\bar{T}$, and $\delta \theta = \theta - \bar{\theta}$.
In addition, for an infinitesimal time interval $dt$, we denote the numbers of adsorption and desorption occurrences by $dN_{\lambda_a}$ and $dN_{\lambda_d}$, respectively, and define $d N_{ad} = dN_{\lambda_a} - dN_{\lambda_d}$.

Before presenting a detailed analysis, we briefly explain the overall approach. 
By introducing a vector $\mathbf{x}^T=\left[\delta\rho,\delta T,\delta T_s,\delta\theta \right]$ to represent the state variables, we express $d\mathbf{x}(t) \equiv \mathbf{x}(t+dt)-\mathbf{x}(t)$ corresponding to the update scheme~\eqref{eq:massenergyupdate} as 
\begin{equation}
\label{eq:xnewz}
    d\mathbf{x} = \mathbf{z}\: d N_{ad}.
\end{equation}
To determine $\mathbf{z}$ that gives a thermodynamically consistent update scheme, we will use the covariance matrix of $\mathbf{x}$, $\mathbf{C} = \left<\mathbf{x}\mathbf{x}^T\right>$, where the brackets denote the equilibrium average.
As can be seen in Eq.~\eqref{eq:lambdaad}, $dN_{ad}$ depends on the instantaneous state $\mathbf{x}$.
By looking at both weak-noise limit and linearized form, we first approximate $dN_{ad}$ as an SDE of the form: 
\begin{equation}
\label{eq:NadSDE}
    d N_{ad} \approx \mathbf{w}_1^T \mathbf{x} dt + w_2\: d W_{ad},
\end{equation}
where $\mathbf{w}_1^T$ and $w_2$ are to be determined below and $W_{ad}$ is a standard Wiener process.
We then obtain a linear SDE for $\mathbf{x}$, see Eq.~\eqref{eq:xSDEAb}, from which we determine $\mathbf{z}$ using a condition that the equilibrium covariance $\mathbf{C}$ should satisfy, see Eq.~\eqref{eq:condC2}.

To obtain an SDE form of $N_{\lambda_a}$, we use the Gaussian approximation of a Poisson process for large $\lambda$: $d N_\lambda \approx \lambda dt + \sqrt{\lambda} dW$, where $W$ denotes a standard Wiener process; note that we assume $N_{tot}$ is sufficiently large.
We also linearize $k_a(T)$, $p(\rho,T)$, and $1-\theta$ terms for small $\delta\rho$, $\delta T$, and $\delta\theta$ around $\bar{\rho}$, $\bar{T}$, and $\bar{\theta}$: 
\begin{subequations}
\begin{align}
    &k_a
    = \bar{k}_a \left\{ 1 +
        \left( \frac{\alpha_a}{k_B\bar{T}}+\beta_{a}\right) \frac{\delta T}{\bar{T}} \right\}, \\
    &p = \bar{p}
        \left\{ 1 + \frac{\delta\rho}{\bar{\rho}}+\frac{\delta T}{\bar{T}}\right\}, \\
    &1-\theta = (1-\bar{\theta})\left\{1 - \frac{\delta\theta}{1-\bar{\theta}}\right\},
\end{align}    
\end{subequations}
where $\bar{k}_a$ and $\bar{p}$ denote corresponding values at $\bar{\rho}$ and $\bar{T}$.
Hence, by using a standard Wiener process $W_a$, we approximate $d N_{\lambda_a}$ as
\begin{equation}
\begin{split}
    d N_{\lambda_a}
    &= \bar{k}_a\bar{p}(1-\bar{\theta})
        \left\{1 + \frac{\delta\rho}{\bar{\rho}} +
            \left( \frac{\alpha_a}{k_B\bar{T}}+\beta_a+1 \right) \frac{\delta T}{\bar{T}}
            -\frac{\delta\theta}{1-\bar{\theta}} \right\} N_{tot}dt \\
    &+ \sqrt{\bar{k}_a\bar{p}(1-\bar{\theta}) N_{tot}}\;d W_a.     
\end{split}    
\end{equation}
Following a similar procedure, we obtain
\begin{equation}
    d N_{\lambda_d}
    = \bar{k}_d\bar{\theta}
        \left\{1 +
            \left( \frac{\alpha_d}{k_B\bar{T}}+\beta_d \right) \frac{\delta T_s}{\bar{T}}
            + \frac{\delta\theta}{\bar{\theta}} \right\} N_{tot}dt
    + \sqrt{\bar{k}_d\bar{\theta} N_{tot}}\;d W_d,     
\end{equation}
where $W_d$ is another standard Wiener process. 
By introducing $\bar{r} \equiv \bar{k}_a\bar{p}(1-\bar{\theta}) = \bar{k}_d\bar{\theta}$ and $W_{ad} = (W_a - W_d)/\sqrt{2}$, we obtain Eq.~\eqref{eq:NadSDE} with
\begin{align}
    &\mathbf{w}_1^T = \bar{r} N_{tot} \left[
        \frac{1}{\bar{\rho}}, 
        \frac{1}{\bar{T}}\left(\frac{\alpha_a}{k_B\bar{T}} + \beta_a + 1\right),
        -\frac{1}{\bar{T}}\left(\frac{\alpha_d}{k_B\bar{T}}+ \beta_d\right),
        -\frac{1}{\bar{\theta}(1-\bar{\theta})} \right], \\
    &w_2 = \sqrt{2\bar{r}N_{tot}}.
\end{align}

By combining Eqs.~\eqref{eq:xnewz} and \eqref{eq:NadSDE}, we obtain an SDE for $\mathbf{x}$:
\begin{equation}
\label{eq:xSDEAb}
    d\mathbf{x} = \mathbf{A}\mathbf{x}\: dt + \mathbf{b}\: d W_{ad},
\end{equation}
where
\begin{equation}
\label{eq:specialAb}
    \mathbf{A} = \mathbf{z}\mathbf{w}_1^T,\quad \mathbf{b} = w_2 \mathbf{z}.
\end{equation}
Eq.~\eqref{eq:xSDEAb} is an Ornstein--Uhlenbeck process whose covariance $\mathbf{C}$ is given by
\begin{equation}
\label{eq:condC2}
    \mathbf{C}\mathbf{A}^T + \mathbf{A}\mathbf{C} + \mathbf{b}\mathbf{b}^T = 0.
\end{equation}
With our choice of state variables, $\mathbf{C}$ is a diagonal matrix~\cite{BalakrishnanGarciaDonevBell2014, Hill1987, Pathria1996},
\begin{equation}
    \mathbf{C} = \mathrm{diag}\left[
        \frac{m}{V}\bar{\rho},
        \frac{k_B\bar{T}^2}{\bar{C}},
        \frac{k_B\bar{T}^2}{\bar{C}_s},
        \frac{\bar{\theta}(1-\bar{\theta})}{N_{tot}}
    \right],
\end{equation}
where $\bar{C}$ and $\bar{C}_s$ are the heat capacities of the gas and solid subsystems at equilibrium, respectively.
Note that while heat capacities $C$ and $C_s$ generally depend on $\rho$ and $\theta$, they can be replaced by their mean values $\bar{C}$ and $\bar{C}_s$ in the weak-noise linearized stochastic analysis.

For a diagonal matrix $\mathbf{C}$, it can be shown that Eqs.~\eqref{eq:specialAb} and \eqref{eq:condC2} have a unique nonzero vector solution:
\begin{equation}
\label{eq:solz}
    \mathbf{z} = -\frac{2}{w_2^2}\mathbf{C}\mathbf{w}_1.
\end{equation}
Thus we obtain
\begin{equation}
    \mathbf{z} = \left[
        -\frac{m}{V},
        -\frac{k_B\bar{T}}{\bar{C}}\left( \frac{\alpha_a}{k_B\bar{T}}+\beta_a+1 \right),
        \frac{k_B\bar{T}}{\bar{C}_s}\left( \frac{\alpha_d}{k_B\bar{T}}+\beta_d \right),
        \frac{1}{N_{tot}}
    \right].
\end{equation}
Therefore, we recover the mass--energy update scheme~\eqref{eq:massenergyupdate} with
\begin{equation}
    \sigma q = -\alpha_a - (\beta_a + 1 ) k_B\bar{T}, \quad
    (1-\sigma) q = \alpha_d + \beta_d k_B\bar{T},
\end{equation}
and heat of adsorption given
\begin{equation}
    q = (\alpha_d-\alpha_a) + (\beta_d-\beta_a-1) k_B\bar{T}.
\end{equation}
By using the TCR model results, Eqs.~\eqref{eq:alphabetarel}--\eqref{eq:eads} and \eqref{eq:alphaKbetaK}, we obtain
\begin{equation}
    q = m \left\{e_g(\bar{T}) - e_{ads}(\bar{T})\right\}.
\end{equation}
Recall that we have dropped the species index $\chem{A}$ in $\rho\equiv \rho_\chem{A}$, $p\equiv p_\chem{A}$, etc.\ so far for notational simplicity and $e_g(\bar{T})\equiv e_{g,\chem{A}}(\bar{T})$ is the mean specific internal energy of gas species $\chem{A}$.

\subsection{\label{sec:PhysicalInterpretation}Physical Interpretation}

For the physical modeling setting described by the full hydrodynamic system in Section~\ref{sec:FullSystemDescription}, we give a physical interpretation of our mass--energy update.
For the TCR model, we choose the $\alpha_a$ and $\beta_a$ parameters of $k_a$ and use Eq.~\eqref{eq:relKkakd} to determine $k_d$.
We assume that the mean adsorption rate is proportional to the mean collision rate:
\begin{equation}
    \bar{r}_a = f \bar{r}_{col},
\end{equation}
where $f$ is the sticking coefficient~\cite{ReuterScheffler2006}.
Since $\bar{r}_{col}$ is given as
\begin{equation}
    \bar{r}_{col} = \frac{\mathcal{A}p}{\sqrt{2\pi m k_B T}}
\end{equation}
for an area of $\mathcal{A}$, we have from Eq.~\eqref{eq:modifiedArrhenius}
\begin{equation}
\label{eq:alphaabetaa}
    \alpha_a = 0, \quad \beta_a = -\frac{1}{2}.
\end{equation}
In addition, to simplify our physical modeling of the solid subsystem, we assume that the temperature of the surface is fixed to $T_s = \bar{T}$, equivalently, $\langle\delta T_s^2\rangle = k_B \bar{T}^2/\bar{C}_s\rightarrow 0$, by taking the infinite heat capacity limit $\bar{C}_s\rightarrow\infty$.

\begin{figure}
\includegraphics[width=\linewidth]{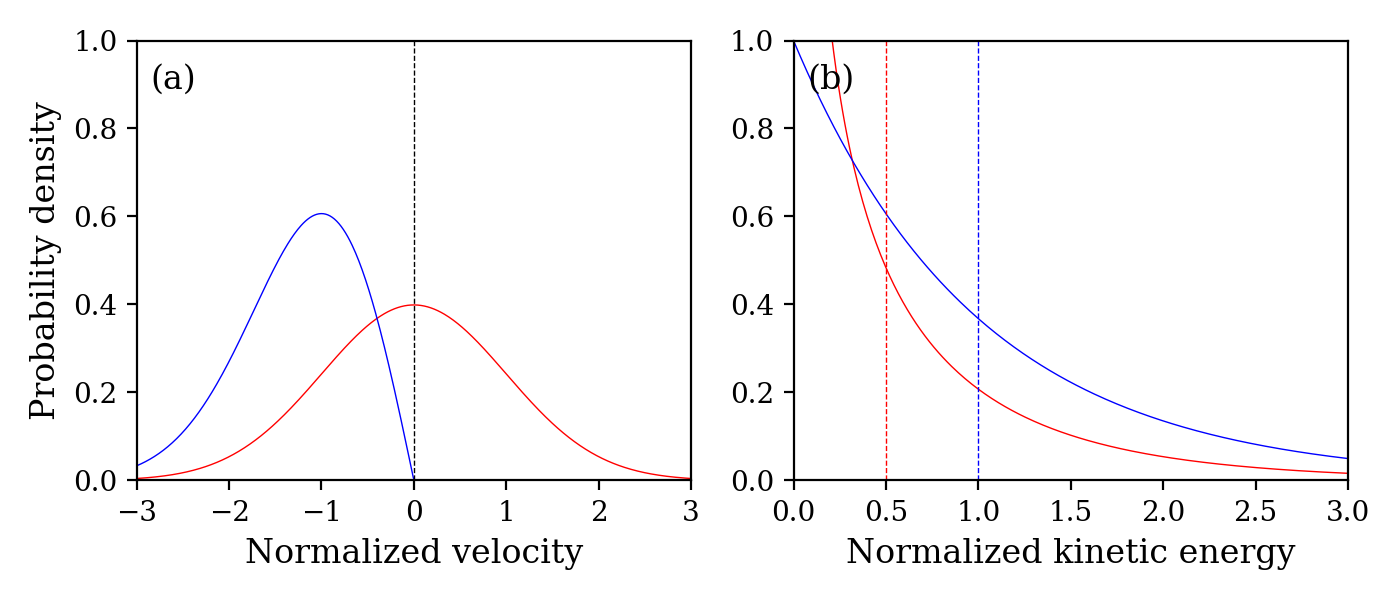}
    \caption{\label{Fig:PDFMaxwellBoltzmannRayleigh}
    For a gas molecule colliding with a wall normal to the $z$ axis (blue) and a gas molecule far from the wall (red), the probability density functions of (a) the normal velocity component, $v_z$, and (b) the corresponding kinetic energy, $\frac12 m v_z^2$, are compared. 
    In panel (a), velocity is normalized by $\sqrt{k_B T/m}$.
    The blue curve depicts the Rayleigh distribution, whereas the red curve shows the Maxwell--Boltzmann distribution.
    In panel (b), kinetic energy is normalized by $k_B T$.
    The vertical dashed lines indicate the mean kinetic energies: $k_B T$ for the Rayleigh distribution (blue) and $\frac12 k_B T$ for the Maxwell--Boltzmann distribution (red).} 
\end{figure}

We investigate the change in the total energy density $\mathcal{E}$ of the gas subsystem due to the adsorption-desorption count $d N_{ad} = d N_{\lambda_a} - d N_{\lambda_d}$.
Since $\mathcal{E} = \rho_\chem{A} e_{g,\chem{A}}(T) + \rho_\chem{B} e_{g,\chem{B}}(T)$ is expressed as
\begin{equation}
    \mathcal{E} 
    = \bar{\mathcal{E}} + e_{g,\chem{A}}(\bar{T})\delta\rho_\chem{A} + \frac{\bar{C}}{V}\delta T,
\end{equation}
the energy density change due to reversible adsorption is given as
\begin{equation}
\label{eq:dmathcalE}
    d\mathcal{E} 
    = -\frac{1}{V} \Bigl\{ m_\chem{A} e_{g,\chem{A}}(\bar{T}) -\sigma q \Bigr\} d N_{ad}
    = -\frac{1}{V} \Bigl\{ m_\chem{A} e_{g,\chem{A}}(\bar{T}) +\frac12 k_B \bar{T} \Bigr\} d N_{ad}
\end{equation}
This implies that, for an adsorption event (or desorption event), the energy of the gas subsystem should be decreased (or increased) by $m_\chem{A} e_{g,\chem{A}}(\bar{T}) +\frac12 k_B \bar{T}$.
The additional energy term, $-\sigma q = \frac12 k_B \bar{T}$, is attributed to the difference in the mean kinetic energy of a gas molecule colliding with the wall
\begin{equation}
    \frac{m}{2}\left<v_x^2+v_y^2+v_z^2\right> 
    = \frac12 k_B \bar{T} + \frac12 k_B \bar{T} + k_B \bar{T} = 2 k_B \bar{T}
\end{equation}
from the value obtained from the Maxwell--Boltzmann distribution, $\frac32 k_B \bar{T}$.
In other words, the normal velocity component of colliding gas molecules has a Rayleigh distribution; that component has an average kinetic energy of $k_B \bar{T}$ while it is $\frac12 k_B \bar{T}$ for each of the other two velocity components, see Figure~\ref{Fig:PDFMaxwellBoltzmannRayleigh}.

\section{\label{sec:FullSystemDescription}Full-System Description}

In this section, we construct a thermodynamically consistent FHD formulation for the full hydrodynamic system of a gas-solid interface undergoing reversible adsorption.
To this end, we first consider an FHD formulation for the corresponding gas-solid interfacial system \textit{without} reversible adsorption and then incorporate the Langmuir adsorption into the FHD formulation.
More specifically, we assume that the non-adsorption FHD formulation is thermodynamically consistent and confirm that embedding reversible adsorption preserves the established thermodynamic equilibrium.
Since we need gas cells contacting the surface for the construction of a coupling between the gas dynamics and the surface coverage dynamics, instead of continuous-space FHD description formally given as stochastic partial differential equations (SPDEs), we use a spatially discretized version of FHD.

To simplify physical modeling, we assume that the surface is connected to an infinite heat bath and thus the surface temperature remains constant (i.e., $T_s \equiv \bar{T}$) and the desorption rate constant $k_d$ does not vary, i.e., $\bar{k}_d \equiv k_d(\bar{T})$. 
In addition, as mentioned for the temperature dependence of the adsorption rate constant $k_a$ in Section~\ref{sec:PhysicalInterpretation}, we assume that the adsorption rate is proportional to the gas-surface collision rate (i.e., $\alpha_a = 0$ and $\beta_a = -\frac{1}{2}$). 
As in Section~\ref{sec:MassEnergyUpdate}, for simplicity of the exposition, we consider an ideal gas mixture of a chemical species $\chem{A}$ undergoing reversible adsorption and a nonreactive species~$\chem{B}$.
In Section~\ref{sec:FHDwithoutads}, we introduce a spatially discretized FHD description of a gas-solid interfacial system without adsorption as a starting point of our construction.
In Section~\ref{sec:EmbeddingLangmuirAdsoprtion}, we construct a thermodynamically consistent coupling between the gas dynamics and the surface coverage dynamics.

\subsection{\label{sec:FHDwithoutads}Spatially Discretized FHD Description without Adsorption}

Since our approach to incorporate the Langmuir adsorption is applicable to a general class of continuum-based mesoscopic simulation methods, we will assume a general form of the time evolution equations, which are linearized for a spatially discretized FHD model, as a starting point of our construction, see Eq.~\eqref{eq:OUfordeltaQ}.
However, for completeness of the exposition, we first explain how this form is obtained.
For a two-species gas system, the time evolution of the conservative variables, $\mathbfcal{U}(\boldsymbol{x},t) = \left[\rho_\chem{A}(\boldsymbol{x},t), \rho_\chem{B}(\boldsymbol{x},t), \rho\boldsymbol{v}(\boldsymbol{x},t), \rho E(\boldsymbol{x},t)\right]^T$ (species mass densities of $\chem{A}$ and $\chem{B}$, momentum density, and energy density, respectively, with the total mass density being $\rho = \rho_\chem{A} + \rho_\chem{B}$), is given formally as SPDEs of the following form:
\begin{equation}
\label{eq:2specFHDformal}
    \frac{\partial \mathbfcal{U}}{\partial t}  
    = -\nabla\cdot\Bigl[
        \mathbf{F}_H(\mathbfcal{Q}) + \mathbf{F}_D(\mathbfcal{Q}) + \mathbf{F}_S(\mathbfcal{Q})
    \Bigr]
    + \mathbf{S}(\mathbfcal{Q}),
\end{equation}
where $\mathbf{F}_H$, $\mathbf{F}_D$, and $\mathbf{F}_S$ are the hyperbolic, diffusive, and stochastic fluxes and $\mathbf{S}$ represents source terms (e.g., reactions, gravitational body forces).
Note that it is more convenient to express the terms in the right-hand side of Eq.~\eqref{eq:2specFHDformal} as functions of the primitive variables, $\mathbfcal{Q}(\boldsymbol{x},t) = \left[\rho_\chem{A}(\boldsymbol{x},t), \rho_\chem{B}(\boldsymbol{x},t), \boldsymbol{v}(\boldsymbol{x},t), T(\boldsymbol{x},t)\right]^T$, which contains the same information as the conservative variables $\mathbfcal{U}(\boldsymbol{x},t)$.
The relation between the total specific energy $E$ and the temperature $T$ is given in Eq.~\eqref{eq:Eexpression}.
For a more detailed description of the SPDEs, see Appendix~\ref{sec:appendixFHDreactgasmix}.
For the physical boundary where the Langmuir adsorption model is to be embedded, which we assume to be located at plane $z=0$, we impose boundary conditions corresponding to an impermeable wall~\cite{SrivastavaLadigesNonakaGarciaBell2023}.

\begin{figure}
\includegraphics[width=0.8\linewidth]{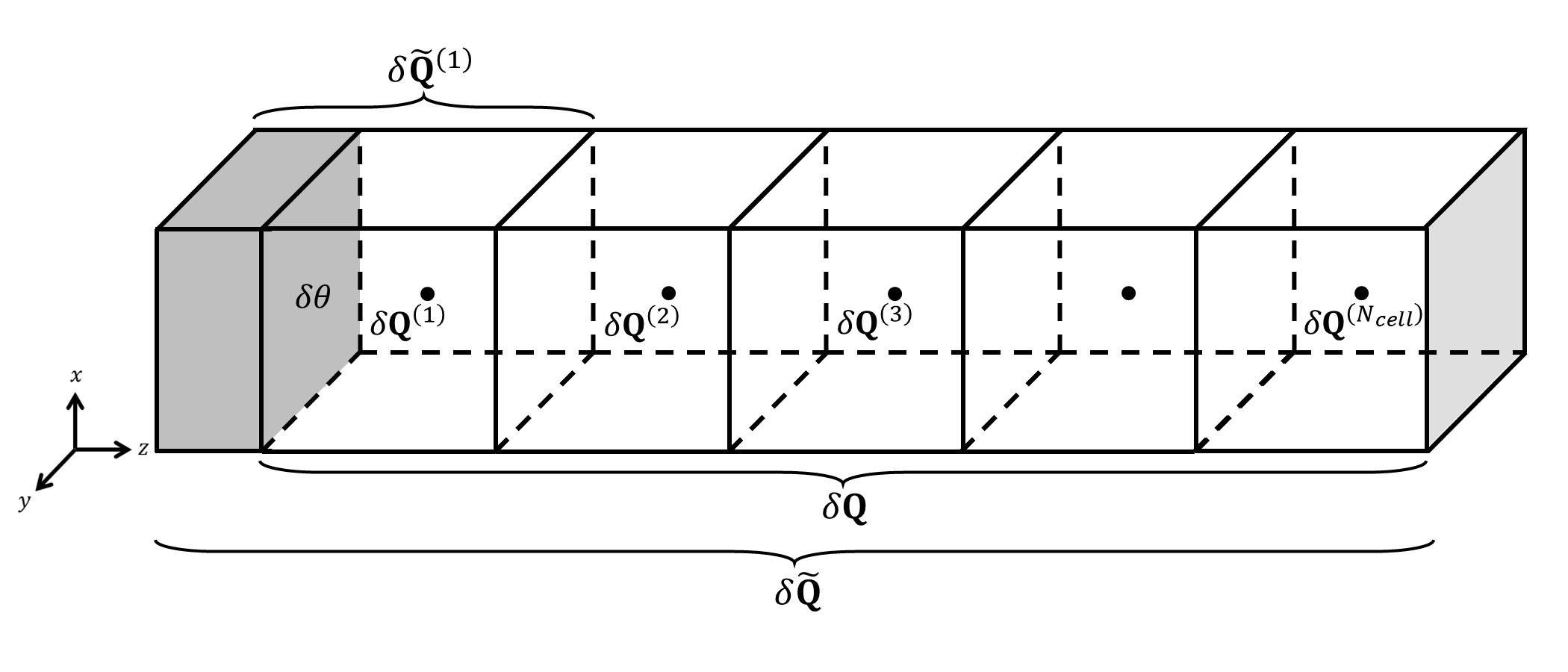}
    \caption{\label{Fig:SpatiallyDiscretizedDescription}
    An illustration of the spatially discretized system.
    It consists of $N_{cell}$ gas cells and an adsorbent surface.
    The states of gas cells are described by $\delta\mathbf{Q}^{(i)}$, $i=1,2,\dots,N_{cell}$ (and collectively by $\delta\mathbf{Q}$), and the state of the surface is described by $\delta\theta$.
    It is also shown which variables the augmented variables $\delta\tilde{\mathbf{Q}}^{(1)}$ and $\delta\tilde{\mathbf{Q}}$ contain.}
\end{figure}

To analytically investigate the thermodynamic consistency condition, we consider the weak-noise limit and linearize Eq.~\eqref{eq:2specFHDformal} around the equilibrium state to obtain linear SPDEs of $\delta\mathbfcal{U}(\boldsymbol{x},t) = \mathbfcal{U}(\boldsymbol{x},t) - \bar{\mathbfcal{U}}$.
However, since our adsorption model is more easily described in terms of primitive variables (i.e., temperature rather than energy density), we consider equivalent linear SPDEs for $\delta\mathbfcal{Q}(\boldsymbol{x},t)=\mathbfcal{Q}(\boldsymbol{x},t) - \bar{\mathbfcal{Q}}$, which can be expressed as
\begin{equation}
\label{eq:linearSPDEforQ}
    d(\delta\mathbfcal{Q}) 
    = \mathbfcal{A}\:\delta\mathbfcal{Q}\:dt
    + \mathbfcal{B}\:d\mathbfcal{W}.
\end{equation}
Here, $\mathbfcal{A}$ and $\mathbfcal{B}$ are linear operators acting on $\delta\mathbfcal{Q}$ and a cylindrical Wiener process (Brownian sheet) $\mathbfcal{W}$, respectively.
We then spatially discretize these linear SPDEs~\eqref{eq:linearSPDEforQ} assuming that all primitive variables are located at cell centers.
We express the resulting system of stochastic ordinary differential equations as
\begin{equation}
\label{eq:OUfordeltaQ}
    d(\delta\mathbf{Q}) = \mathbf{A}_{hyd}\:\delta\mathbf{Q}\:dt + \mathbf{B}_{hyd}\:d\mathbf{W},
\end{equation}
where $\delta\mathbf{Q}$, $\mathbf{W}$, $\mathbf{A}_{hyd}$, and $\mathbf{B}_{hyd}$ represent spatial discretizations of $\delta\mathbfcal{Q}$, $\mathbfcal{W}$, $\mathbfcal{A}$, and $\mathbfcal{B}$, respectively.
Note that the subscript $_{hyd}$ is used in $\mathbf{A}_{hyd}$, and $\mathbf{B}_{hyd}$ to emphasize that these matrices are for the hydrodynamic update (as opposed to the reversible adsorption update, which will be introduced in Section~\ref{sec:EmbeddingLangmuirAdsoprtion}). 
For simplicity of exposition, we further assume a one-dimensional system consisting of $N_{cell}$ gas cells, see Figure~\ref{Fig:SpatiallyDiscretizedDescription}.
In this case, the state of the overall system is represented by
\begin{equation}
    \delta\mathbf{Q}(t) = 
    \begin{bmatrix}
        \delta\mathbf{Q}^{(1)}(t) \\
        \delta\mathbf{Q}^{(2)}(t) \\
        \vdots \\
        \delta\mathbf{Q}^{(N_{cell})}(t)
    \end{bmatrix},\;\;
    \mbox{where}\;\;
    \delta\mathbf{Q}^{(i)}(t) =
    \begin{bmatrix}
        \delta\rho_\chem{A}^{(i)}(t) \\
        \delta\rho_\chem{B}^{(i)}(t) \\
        \delta v_z^{(i)}(t) \\
        \delta T^{(i)}(t)
    \end{bmatrix}
\end{equation}
denotes the state of cell $i$.

From the theory of stochastic processes~\cite{Gardiner2004, DonevVanden-EijdenGarciaBell2010}, the correlation matrix
$\mathbf{C}_\mathbf{Q} = \langle \delta \mathbf{Q} \: \delta \mathbf{Q}^T \rangle$
for the Ornstein--Uhlenbeck process~\eqref{eq:OUfordeltaQ} satisfies
\begin{equation}
\label{eq:condCQ}
    \mathbf{A}_{hyd} \mathbf{C}_\mathbf{Q} + \mathbf{C}_\mathbf{Q} \mathbf{A}_{hyd}^T
    + \mathbf{B}_{hyd} \mathbf{B}_{hyd}^T = 0.
\end{equation}
Since we assume that the non-adsorption FHD formulation is thermodynamically consistent, $\mathbf{C}_\mathbf{Q}$ is a diagonal matrix where each diagonal component can be given by equilibrium statistical mechanics~\cite{BalakrishnanGarciaDonevBell2014}:
\begin{equation}
\label{eq:equilfluct1}
    \langle\delta\rho_A^2 \rangle = \frac{m_\chem{A}\bar{\rho}_\chem{A}}{\Delta V},\;
    \langle\delta\rho_B^2 \rangle = \frac{m_\chem{B}\bar{\rho}_\chem{B}}{\Delta V},\;
    \langle\delta v_z^2 \rangle 
    = \frac{k_B \bar{T}}{\Delta V (\bar{\rho}_\chem{A} + \bar{\rho}_\chem{B})},\;
    \langle\delta T^2 \rangle
    = \frac{k_B \bar{T}^2}{\Delta V (c_{v,\chem{A}}\bar{\rho}_\chem{A} + c_{v,\chem{B}}\bar{\rho}_\chem{B})},
\end{equation}
where $\Delta V$ is the volume of each gas cell.

\subsection{\label{sec:EmbeddingLangmuirAdsoprtion}Embedding Langmuir Adsorption}

Our next step is to augment the gas system (described by $\delta\mathbf{Q}$) to incorporate the surface coverage (described by $\delta\theta$), see Figure~\ref{Fig:SpatiallyDiscretizedDescription}, by defining $\delta\tilde{\mathbf{Q}}$ as
\begin{equation}
\label{eq:deltilQrep}
    \delta\tilde{\mathbf{Q}}(t) =
    \begin{bmatrix}
        \delta\theta(t) \\
        \delta\mathbf{Q}(t)
    \end{bmatrix} =
    \begin{bmatrix}
        \delta\tilde{\mathbf{Q}}^{(1)}(t) \\
        \delta\mathbf{Q}^\mathrm{bulk}(t)
    \end{bmatrix},
\end{equation}
where
\begin{equation}
    \delta\tilde{\mathbf{Q}}^{(1)}(t) =
    \begin{bmatrix}
        \delta\theta(t) \\
        \delta\mathbf{Q}^{(1)}(t)
    \end{bmatrix}, \quad
    \delta\mathbf{Q}^\mathrm{bulk}(t) = 
    \begin{bmatrix}
        \delta\mathbf{Q}^{(2)}(t) \\
        \vdots \\
        \delta\mathbf{Q}^{(N_{cell})}(t)
    \end{bmatrix}.
\end{equation}
Note that both representations of $\delta\tilde{\mathbf{Q}}$ given in Eq.~\eqref{eq:deltilQrep} are needed to describe the time evolution of the overall system.
More specifically, hydrodynamics updates the state variables of all gas cells (represented by $\delta\mathbf{Q}$), whereas reversible adsorption updates the state variables of the first cell as well as the surface coverage (i.e., $\delta\tilde{\mathbf{Q}}^{(1)}$). 
Note that the state variables of the first cell (i.e., $\delta\mathbf{Q}^{(1)}$) are included in both $\delta\mathbf{Q}$ and $\delta\tilde{\mathbf{Q}}^{(1)}$ and are updated by both hydrodynamics and reversible adsorption.

We now follow similar steps to Section~\ref{sec:MassEnergyUpdate} to obtain an SDE (see Eq.~\eqref{eq:dadstilQ1}) for the change in $\delta\tilde{\mathbf{Q}}^{(1)}$ due to reversible adsorption.
As in Eq.~\eqref{eq:xnewz}, we assume that this change has the form
\begin{equation}
    d_{ads}(\delta\tilde{\mathbf{Q}}^{(1)}) = \mathbf{z}\;dN_{ad},
\end{equation}
where $d_{ads}$ denotes the change due to reversible adsorption.
Here, $\mathbf{z}$ is to be determined and the adsorption-desorption count is given as $dN_{ad} = d N_{\lambda_a} - d N_{\lambda_d}$, where
\begin{equation}
\label{eq:lambdaadFHD}
    \lambda_a = k_a(T^{(1)})p\bigl(\rho_\chem{A}^{(1)},T^{(1)}\bigr)(1-\theta)N_{tot},\quad
    \lambda_d = \bar{k}_d\theta N_{tot}.
\end{equation}
Recall that $N_{tot}$ is the total number of adsorption sites of cell~1 and that the surface temperature as fixed so $\bar{k}_d = k_d \left(\bar{T}\right)$.
By linearizing $d N_{ad}$ around the equilibrium state in the weak-noise limit, we obtain the approximation
\begin{equation}
    d N_{ad} \approx \mathbf{w}_1^T \delta\tilde{\mathbf{Q}}^{(1)} dt + w_2\: d W_{ad},
\end{equation}
where
\begin{equation}
    \mathbf{w}_1^T = \bar{r} N_{tot} \left[
        -\frac{1}{\bar{\theta}(1-\bar{\theta})},
        \frac{1}{\bar{\rho}_\chem{A}},
        0,
        0,
        \frac{1}{2\bar{T}}
    \right], \quad
    w_2 = \sqrt{2\bar{r}N_{tot}}
\end{equation}
and $\bar{r} = \bar{k}_a \bar{p}(1-\bar{\theta}) = \bar{k}_d\bar{\theta}$.
This allows us to write
\begin{equation}
\label{eq:dadstilQ1}
    d_{ads}(\delta\tilde{\mathbf{Q}}^{(1)}) 
    = \mathbf{z}\:\mathbf{w}_1^T\:\delta\tilde{\mathbf{Q}}^{(1)}\:dt 
    + w_2\:\mathbf{z}\:d W_{ad}
    \equiv \mathbf{A}_{ads}\:\delta\tilde{\mathbf{Q}}^{(1)}\:dt
    + \mathbf{b}_{ads}\:d W_{ad}.
\end{equation}
Hence, the time evolution of the overall system is given as
\begin{equation}
\label{eq:ddeltatilQ}
    d (\delta\tilde{\mathbf{Q}}) =
    \begin{bmatrix}
        0 \\ \mathbf{A}_{hyd}\:\delta\mathbf{Q}\:dt + \mathbf{B}_{hyd}\:d\mathbf{W}
    \end{bmatrix} +
    \begin{bmatrix}
        \mathbf{A}_{ads} \delta\tilde{\mathbf{Q}}^{(1)} dt + \mathbf{b}_{ads} \: d W_{ad} \\
        \mathbf{0}
    \end{bmatrix}.
\end{equation}
Note that the first and second terms on the right-hand side of Eq.~\eqref{eq:ddeltatilQ} are based on the first and second representations of $\delta\tilde{\mathbf{Q}}$ in Eq.~\eqref{eq:deltilQrep}, respectively.
Eq.~\eqref{eq:ddeltatilQ} can be written as
\begin{equation}
\label{eq:timeevoldeltilQ}
    d (\delta\tilde{\mathbf{Q}}) 
    = (\tilde{\mathbf{A}}_{hyd} + \tilde{\mathbf{A}}_{ads})\:\delta\tilde{\mathbf{Q}}
    + \tilde{\mathbf{B}}_{hyd}\:d\mathbf{W}
    + \tilde{\mathbf{b}}_{ads}\:dW_{ad}
\end{equation}
by defining the matrices
\begin{equation}
    \tilde{\mathbf{A}}_{hyd} =
    \begin{bmatrix}
        0 & \mathbf{0} \\
        \mathbf{0} & \mathbf{A}_{hyd}
    \end{bmatrix},\;
    \tilde{\mathbf{A}}_{ads} =
    \begin{bmatrix}
        \mathbf{A}_{ads} & \mathbf{0} \\
        \mathbf{0} & \mathbf{0}
    \end{bmatrix},\;
    \tilde{\mathbf{B}}_{hyd} =
    \begin{bmatrix}
        \mathbf{0} \\
        \mathbf{B}_{hyd}
    \end{bmatrix},\;
    \tilde{\mathbf{b}}_{ads} =
        \begin{bmatrix}
        \mathbf{b}_{ads} \\
        \mathbf{0}
    \end{bmatrix}.
\end{equation}
As before, the correlation matrix $\mathbf{C}_{\tilde{\mathbf{Q}}} = \langle \delta \tilde{\mathbf{Q}} \: \delta \tilde{\mathbf{Q}}^T \rangle$ for this Ornstein--Uhlenbeck process is given by
\begin{equation}
\label{eq:condCQtilde}
    (\tilde{\mathbf{A}}_{hyd} + \tilde{\mathbf{A}}_{ads})\mathbf{C}_{\tilde{\mathbf{Q}}}
    + \mathbf{C}_{\tilde{\mathbf{Q}}} (\tilde{\mathbf{A}}_{hyd} + \tilde{\mathbf{A}}_{ads})^T
    + \tilde{\mathbf{B}}_{hyd}\tilde{\mathbf{B}}_{hyd}^T
    + \tilde{\mathbf{b}}_{ads} \tilde{\mathbf{b}}_{ads}^T
    = \mathbf{0}.
\end{equation}
Since we want the overall time evolution~\eqref{eq:timeevoldeltilQ} to reproduce thermodynamic equilibrium, we require that $\mathbf{C}_{\tilde{\mathbf{Q}}}$ is given by a diagonal matrix where the first diagonal component is given as
\begin{equation}
\label{eq:equilfluct2}
    \langle\delta\theta^2\rangle = \frac{\bar{\theta}(1-\bar{\theta})}{N_{tot}}
\end{equation}
and the other diagonal components are given from $\mathbf{C}_\mathbf{Q}$.

By comparing Eq.~\eqref{eq:condCQ} (for $\mathbf{C}_\mathbf{Q}$) and Eq.~\eqref{eq:condCQtilde} (for $\mathbf{C}_{\tilde{\mathbf{Q}}}$), we obtain the following equation for $\mathbf{C}_{\tilde{\mathbf{Q}}^{(1)}} = \langle \delta \tilde{\mathbf{Q}}^{(1)} \: (\delta \tilde{\mathbf{Q}}^{(1)})^T \rangle$:
\begin{equation}
\label{eq:condCtildeQ1}
    \mathbf{A}_{ads}\mathbf{C}_{\tilde{\mathbf{Q}}^{(1)}}
    + \mathbf{C}_{\tilde{\mathbf{Q}}^{(1)}}^T \mathbf{A}_{ads}
    + \mathbf{b}_{ads} \mathbf{b}_{ads}^T
    = \mathbf{0}.
\end{equation}
Note that the dimensions of $\mathbf{C}_\mathbf{Q}$, $\mathbf{C}_{\tilde{\mathbf{Q}}}$, $\mathbf{C}_{\tilde{\mathbf{Q}}^{(1)}}$ are $4 N_{cell}\times 4 N_{cell}$, $(4 N_{cell}+1)\times (4 N_{cell}+1)$, $5 \times 5$, respectively.
We first extend Eq.~\eqref{eq:condCQ} to the $(4 N_{cell}+1)\times (4 N_{cell}+1)$ space to subtract it from Eq.~\eqref{eq:condCQtilde} and then reduce the resulting equation to the $5 \times 5$ space to obtain Eq.~\eqref{eq:condCtildeQ1}.

Finally, as we did in Section~\ref{sec:MassEnergyUpdate}, we determine the unknown vector $\mathbf{z}$ by solving Eq.~\eqref{eq:condCtildeQ1} with the thermodynamic equilibrium values of $\mathbf{C}_{\tilde{\mathbf{Q}}^{(1)}}$:
\begin{equation}
    \mathbf{z} = -\frac{2}{w_2^2}\mathbf{C}_{\tilde{\mathbf{Q}}^{(1)}} \mathbf{w}_1 =
    \left[
        \frac{1}{N_{tot}}, \;
        -\frac{m_\chem{A}}{\Delta V}, \;
        0, \;
        0, \;
        -\frac{k_B \bar{T}}{2 \Delta V (c_{v,\chem{A}}\bar{\rho}_\chem{A} + c_{v,\chem{B}}\bar{\rho}_\chem{B})}
    \right]^T.
\end{equation}
Hence, for cell~1, the changes in the surface coverage, mass density of \chem{A}, and temperature due to the adsorption-desorption count $dN_{ad} = d N_{\lambda_a} - d N_{\lambda_d}$ are given as
\begin{subequations}
\label{eq:dadsres1}
\begin{align}
    d_{ads}(\theta) &= \frac{1}{N_{tot}} d N_{ad},\\
    d_{ads}(\rho_\chem{A}^{(1)}) &= -\frac{m_\chem{A}}{\Delta V} d N_{ad},\\
    d_{ads}(T^{(1)}) &= -\frac{k_B \bar{T}}{2 \Delta V (c_{v,\chem{A}}\bar{\rho}_\chem{A} + c_{v,\chem{B}}\bar{\rho}_\chem{B})} d N_{ad},
\end{align}    
\end{subequations}
and there is no change in the mass density of \chem{B} or in the normal velocity due to the adsorption-desorption count, that is,
\begin{equation}
\label{eq:dadsres2}
    d_{ads}(\rho_\chem{B}^{(1)}) = 0,\quad
    d_{ads}(v_z^{(1)}) = 0.        
\end{equation}
Using these results, we obtain the change in the total energy density $\mathcal{E}^{(1)}=\rho^{(1)}E^{(1)}$ of cell~1 due to the adsorption-desorption count:
\begin{equation}
\label{eq:dadsmathcalE}
    d_{ads}(\mathcal{E}^{(1)}) = -\frac{1}{\Delta V}\left\{m_\chem{A}e_\chem{A}(\bar{T})+\frac12 k_B \bar{T}\right\} d N_{ad}.
\end{equation}
As discussed in Section~\ref{sec:PhysicalInterpretation}, the energy correction term $\frac12 k_B T$ appears due to the difference in the mean kinetic energy between the Maxwell--Boltzmann distribution and molecules colliding with the surface.
For the normal momentum density $J_z^{(1)}=\rho^{(1)}v_z^{(1)}$ of cell~1, we obtain
\begin{equation}
\label{eq:dadsJz}
    d_{ads}(J_z^{(1)}) = \bar{\rho} \: d_{ads}(v_z^{(1)}) = 0.
\end{equation}
Note that these results are valid up to first order in the weak-noise limit.
Since $\bar{v}_z^{(1)}=0$ in equilibrium and thus $v_z^{(1)} = \delta v_z^{(1)}$, second-order terms like $\frac12\bar{\rho}^{(1)}(\delta v_z^{(1)})^2$ and $\delta\rho^{(1)}\delta v_z^{(1)}$ do not contribute to the final results in Eqs.~\eqref{eq:dadsmathcalE} and \eqref{eq:dadsJz}.

\section{\label{sec:NumericalValidation}Numerical Validation}

In this section, we present a numerical validation study to demonstrate that our numerical method reproduces the correct thermodynamic equilibrium.
To this end, we perform equilibrium simulations of an ideal gas mixture of $\chem{CO}$ and $\chem{Ar}$, where $\chem{CO}$ undergoes reversible adsorption onto an adsorbent surface.
To confirm the thermodynamic equilibrium, we mainly analyze the cell variances and structure factors for the dynamical variables. 
Before presenting our simulation results, we briefly explain the construction and implementation of our numerical method in Section~\ref{subsec:NumMethod} and describe the model system and simulation parameters in Section~\ref{subsec:ModelParams}. 
We present simulation results in Section~\ref{subsec:SimRes}.
We first analyze simulation results obtained by our numerical method in Section~\ref{subsubsec:CorrectScheme}.
To validate our method, we further present simulation results based on alternative methods, which exhibit thermodynamic inconsistency.
In particular, we show  that using mean partial pressure and temperature to evaluate adsorption rate (Section~\ref{subsubsec:MeanScheme}) or omitting the energy correction term from Eq.~\eqref{eq:FHDupdate_c} (Section~\ref{subsubsec:WithoutEnergyUpdateScheme}) both lead to thermodynamically inconsistent results.

\subsection{\label{subsec:NumMethod}Numerical Method}

We construct a numerical method by incorporating our thermodynamically consistent update for reversible adsorption into the compressible FHD solver~\cite{SrivastavaLadigesNonakaGarciaBell2023, PolimenoKimBlanchetteSrivastavaGarciaNonakaBell2025} of the FHDeX software~\cite{FHDeX_github}.
While the reversible adsorption update to be embedded into the FHD solver is essentially the same as the one considered for the analytic stochastic analysis performed in Section~\ref{sec:FullSystemDescription}, see Eqs.~\eqref{eq:dadsres1}--\eqref{eq:dadsJz}, there are a few technically different assumptions that require minor modifications to the setup.
Hence, before presenting our reversible adsorption update and explaining how to couple the Langmuir adsorption model with compressible FHD, we clarify these points.
First, the FHD solver, which is based on the finite-volume approach, solves the time evolution equations of the conservative variables, see Appendix~\ref{sec:appendixFHDreactgasmix}, and uses a staggered grid for momentum density.
Recall that the reversible adsorption update described in the previous section is given in terms of primitive variables and assumes that all variables, including velocity, are located at cell centers.
Second, the FHD solver assumes a three-dimensional domain, whereas a one-dimensional array of gas cells is considered in the analytic stochastic analysis.
We assume that the adsorbent surface is located at the lower wall normal to the $z$ axis (i.e., the plane $z=0$). 
We apply the reversible adsorption update to cells contacting this surface, which have cell indices $\boldsymbol{i} = (i_x,i_y,1)$.
Third, we construct the reversible adsorption update so that it is applicable to nonequilibrium systems beyond the weak-noise limit.  
To this end, the energy of a gas molecule involved in reversible adsorption is computed using the instantaneous temperature $T^{(\boldsymbol{i})}$ of the corresponding cell instead of the equilibrium temperature $\bar{T}$, see Eq.~\eqref{eq:FHDupdate_c}.
Note, however, that this update is reduced to the one considered in the previous section in the weak-noise limit in equilibrium.
In addition, rather than using Gaussian approximation, we use a Poisson random number generator $\mathcal{P}(M)$ with mean $M$ to sample the numbers of adsorption and desorption events, $\Delta N_a$ and $\Delta N_d$.
For a small time interval $\tau$, these numbers are given as
\begin{equation}
\label{eq:NaNdPoisson}
    \Delta N_a = \mathcal{P}(\lambda_a \tau),\quad 
    \Delta N_d = \mathcal{P}(\lambda_d \tau),
\end{equation}
where $\lambda_a$ and $\lambda_d$ are given in Eq.~\eqref{eq:lambdaadFHD}.
We numerically investigate the validity of the update beyond the weak-noise limit in Section~\ref{subsec:SimRes}.

When $\Delta N_a$ adsorption events and $\Delta N_d$ desorption events occur during a small time interval $\tau$ in cell~$\boldsymbol{i}$ contacting the adsorbent surface, based on the results~\eqref{eq:dadsres1}, \eqref{eq:dadsmathcalE}, we update the surface coverage, mass density of $\chem{A}$, and total energy density using
\begin{subequations}
\label{eq:FHDupdate}
\begin{align}
    \Delta_{ads}\:\theta^{(\boldsymbol{i})} &= \frac{1}{N_{tot}}\Delta N_{ad},\\
    \Delta_{ads}\:\rho_\chem{A}^{(\boldsymbol{i})} &= -\frac{m_\chem{A}}{\Delta V}\Delta N_{ad},\\
    \label{eq:FHDupdate_c}
    \Delta_{ads}\:\mathcal{E}^{(\boldsymbol{i})} &= -\frac{1}{\Delta V}\left\{m_\chem{A} e_\chem{A}(T^{(\boldsymbol{i})})+\frac12 k_B T^{(\boldsymbol{i})}\right\} \Delta N_{ad},
\end{align}
\end{subequations}
where $\Delta N_{ad} = \Delta N_a - \Delta N_d$.
Based on the results~\eqref{eq:dadsres2} and \eqref{eq:dadsJz}, we do not update the mass density of $\chem{B}$ or the momentum density for $\Delta N_{ad}$.
Note that momentum density components are located at cell faces (such that the normal components are on faces with the same normal direction) in the staggered-grid discretization and the FHD solver sets the normal momentum density component to be zero on physical boundaries.
Since our adsorption update does not include momentum density and only updates cell-centered variables, there is no essential change in our reversible adsorption update whether the FHD method uses a staggered grid or not.

To couple the Langmuir adsorption model with compressible FHD, we use operator splitting.
In other words, we decompose the time evolution of the overall system into the updates due to non-adsorption hydrodynamics and reversible adsorption and use the FHD solver and the update scheme~\eqref{eq:FHDupdate} to perform these updates.
For numerical accuracy, we employ Strang splitting~\cite{Strang1968}.
For each time step $\Delta t$, the following updates are performed:
\begin{enumerate}[topsep=0pt,noitemsep]
    \item Perform reversible adsorption update for a half time step $\tau = \Delta t /2$. In other words, for each bottom cell contacting the adsorbent surface, sample the adsorption and desorption counts, $\Delta N_a$ and $\Delta N_d$, for $\Delta t/2$, see Eq.~\eqref{eq:NaNdPoisson}.
    Using the update scheme~\eqref{eq:FHDupdate}, update the surface coverage $\theta$, species mass density $\rho_\chem{A}$, and total energy density $\mathcal{E}$ for the adsorption-desorption count $\Delta N_{ad} = \Delta N_a - \Delta N_d$.
    \item Perform non-adsorption hydrodynamics update for the full time step $\Delta t$ using the FHD solver.
    \item Perform another reversible adsorption update for the remaining half time step $\tau = \Delta t /2$ by sampling new $\Delta N_a$ and $\Delta N_d$.
\end{enumerate}
The sampling of variables for statistical averaging occurs at the end of the time step.
We implemented this numerical method as part of the FHDeX software~\cite{FHDeX_github}, which is available at \href{https://github.com/AMReX-FHD/FHDeX.git}{\texttt{https://github.com/AMReX-FHD/FHDeX.git}}.

\subsection{\label{subsec:ModelParams}Model System and Simulation Parameters}

As a model system, we consider an ideal gas mixture of \chem{CO} and \chem{Ar} and assume that \chem{CO} undergoes reversible adsorption onto an adsorbent wall. 
The simulation parameter values are detailed in this section; we use cgs units.
The system domain is a cube with side length $L_x=L_y=L_z=\SI{1.50e-4}{cm}$, which is discretized into $16^3$ cubic cells of side length $\Delta x=\Delta y=\Delta z=\SI{9.36e-6}{cm}$.
We assume that the gas is contained by two parallel walls normal to the $z$ axis at $z=0$ and $z=L_z$ and reversible adsorption occurs on the lower wall at $z=0$.
Except for the mass--energy update applied to the lower wall, we impose the same physical boundary conditions corresponding to impermeable walls held at constant temperature $\bar{T}$.
More specifically, Neumann boundary conditions are applied to impose zero concentration fluxes, Dirichlet boundary conditions are applied to impose constant temperature, and slip (Neumann) conditions are applied to tangential momentum fluxes.
For the $x$ and $y$ directions, periodic boundary conditions are imposed.

\begin{table}[]
    \centering
    \begin{tabular}{|c|c|c|c|}
    \hline
    \multirow{2}{*}{Parameter} & \multirow{2}{*}{Units} & \multicolumn{2}{c|}{Temperature} \\
    \cline{3-4}
     & & \SI{700}{K} & \SI{800}{K} \\
    \hline
    $\varepsilon_\chem{CO}$ & \multirow{2}{*}{\SI{}{erg/g}} & \SI{-4.30e10}{} & \SI{-4.31e10}{} \\ \cline{1-1}\cline{3-4}
    $\varepsilon_\chem{Ar}$ &  & \SI{-9.17e8}{}  & \SI{-9.17e8}{} \\
    \hline
    $c_{v,\chem{CO}}$ & \multirow{2}{*}{\SI{}{erg\cdot g^{-1} K^{-1}}} & \SI{8.16e6}{} & \SI{8.41e6}{} \\ \cline{1-1}\cline{3-4}
    $c_{v,\chem{Ar}}$ &  & \SI{3.12e6}{}  & \SI{3.12e6}{} \\
    \hline
    $\bar{k}_a$ & \SI{}{cm^2 dyn^{-1}{s}^{-1}} & \SI{1.83e2}{} & \SI{1.71e2}{} \\
    \hline
    $\bar{k}_d$ & \SI{}{s^{-1}} & \SI{3.70e7}{} & \SI{1.25e9}{} \\
    \hline
    \end{tabular}
    \caption{Parameter values for the internal energies of gas species \chem{CO} and \chem{Ar} and the adsorption and desorption rate constants used for equilibrium simulations at \SI{700}{K} and \SI{800}{K}.}
    \label{tab:params}
\end{table}

To validate our numerical method using two different pairs of rate constant values, we conduct simulations at two temperature values, $\bar{T}=\SI{700}{K}$ and $\SI{800}{K}$.
For the equal mass fractions (i.e., $\bar{Y}_\chem{CO}=\bar{Y}_\chem{Ar}=0.5$) at pressure $\bar{p}=\SI{1.01e6}{dyn/cm^2}$, the total mass density has $\bar{\rho}=\SI{5.73e-4}{g/cm^3}$ at \SI{700}{K} and $\bar{\rho}=\SI{5.02e-4}{g/cm^3}$ at \SI{800}{K}.
Parameter values for the internal energies of gas species \chem{CO} and \chem{Ar} are determined using the thermochemistry data in the NIST Chemistry WebBook~\cite{LinstromMallard2001}, see Table~\ref{tab:params}.
To evaluate transport coefficients~\cite{HirschfelderCurtissBird1954} (e.g., viscosity), \chem{CO} and \chem{Ar} are assumed to be hard spheres with diameters~\cite{Baker2012} of \SI{3.76e-8}{cm} and \SI{3.40e-8}{cm}, respectively.

We choose parameter values for reversible adsorption based on experimental~\cite{KuhnSzanyiGoodman1992, SuCremerShenSomorjai1997} and simulation~\cite{ReuterScheffler2006, RogalReuterScheffler2008, PiccininStamatakis2014, WangReuter2015, TetenoireJuaristiAlducin2021} studies.
The number of adsorption sites per gas cell contacting the adsorbent surface is chosen to be $N_{tot}=9\times 10^4$ assuming that an adsorption site occupies a square with side length $a_x=a_y=\SI{3.12e-8}{cm}$.
The values of the rate constants, $\bar{k}_a \equiv k_a(\bar{T})$ and $\bar{k}_d \equiv k_d(\bar{T})$, at the equilibrium temperature $\bar{T}$ are shown in Table~\ref{tab:params}.
Note that the magnitudes of the desorption rate constant $\bar{k}_d$ are significantly different at \SI{700}{K} and \SI{800}{K}, which results in significantly different equilibrium surface coverage values: $\bar{\theta}=0.747$ at \SI{700}{K} and $0.076$ at \SI{800}{K}.

We use time step size $\Delta t=10^{-12}\:\SI{}{s}$. 
Each simulation is initiated with the equilibrium values and run for $1.2\times 10^7$ time steps.
To compute equilibrium averages, the first $2\times 10^6$ steps are discarded and the remaining $10^7$ time steps are used to compute the averages.
Note that reversible adsorption is not too fast and the time step size is mainly chosen by considering the computational efficiency and accuracy for non-adsorption FHD.
The characteristic time scale of reversible adsorption, which is estimated as $\bar{r}^{-1} = (\bar{k}_d \bar{\theta})^{-1}$, is less than $10^5$ time steps for both \SI{700}{K} and \SI{800}{K}.
The average number of adsorption (or desorption) events in a bottom cell contacting the adsorbent surface during $\tau=\Delta t/2$ is 1.24 for \SI{700}{K} and 4.25 for \SI{800}{K}.
For some of the results, particularly for Figures~\ref{Fig:CorrectCellVariances} and \ref{Fig:CorrelationCoefficients}, an ensemble of 16 independent simulations were performed to improve the statistical accuracy to convincingly demonstrate that simulation results agree or disagree with theoretical predictions.

\subsection{\label{subsec:SimRes}Simulation Results}

To check whether the thermodynamic equilibrium is correctly reproduced in simulations, we use the fact that the statistical properties of equilibrium fluctuations of thermodynamic variables in a cell can be given by equilibrium statistical mechanics~\cite{Callen1991}.
More specifically, the second moments of the fluctuations in $\rho_\chem{A}$, $\rho_\chem{B}$, $v_x$, $v_y$, $v_z$, $T$, and $\theta$ are given in Eqs.~\eqref{eq:equilfluct1} and \eqref{eq:equilfluct2} and $\left<\delta v_x^2\right> = \left<\delta v_y^2\right> = \left<\delta v_z^2\right>$.
In addition, equilibrium fluctuations in two different cells are uncorrelated, and equilibrium fluctuations of any pair from these variables in the same cell are uncorrelated.
To confirm these, we analyze the following quantities.
First, for thermodynamic variables (denoted by $\phi$), we compute the cell variance.
Since cells with the same $z$ value are equivalent due to peridoic boundary conditions imposed for the $x$ and $y$ directions, we take the average of the cell variance of $\phi$ over those cells to obtain $C_{\phi}(z) = \left<[\phi-\left<\phi\right>_z]^2\right>_z$, where $\left<\cdot\right>_z$ denotes average over the cells belonging to the layer specified by $z$.
In thermodynamic equilibrium, $C_{\phi}(z)$ should be a constant function in $z$ (i.e., $C_\phi(z) \equiv C_{\phi,eq}$). 
Second, to confirm that there are no unphysical correlations among the cells in each layer specified by $z$, we compute the structure factors defined as
\begin{equation}
    S_{\phi}(\mathbf{k}_\perp,z) = \Delta V\:\langle \delta\hat{\phi}(\mathbf{k}_\perp,z)\: \delta\hat{\phi}^*(\mathbf{k}_\perp,z) \rangle
\end{equation}
where $\mathbf{k}_\perp=(k_x,k_y)$ is a wave vector perpendicular to the $z$-axis, $\delta\hat{\phi}(\mathbf{k}_\perp,z)$ is the discrete Fourier transform of the fluctuation $\delta\phi = \phi - \left<\phi\right>_z$ at $z$ for $\mathbf{k}_\perp$, and $\delta\hat{\phi}^*(\mathbf{k}_\perp,z)$ is its complex conjugate.
At thermodynamic equilibrium, the structure factor spectra become flat (i.e., constant functions in $\mathbf{k}_\perp$) with the value
\begin{equation}
    S_{\phi,eq} = \Delta V\: C_{\phi,eq}.
\end{equation}
Third, we compute the correlation coefficients of equilibrium fluctuations 
\begin{equation}
\label{eq:corrcoeff}
    r_{\phi,\phi'} = \frac{\bigl<\bigl(\phi-\left<\phi\right>\bigr)\left(\phi'-\left<\phi'\right>\right)\bigr>}{\sqrt{C_\phi}\sqrt{C_{\phi'}}},
\end{equation}
for pairs of variables in a bottom cell contacting the adsorbent surface, including $(\phi,\phi')=(\rho_\chem{CO},T)$, $(\theta,T)$, and $(\theta,\rho_\chem{CO})$.
Note that we drop the subscript $z$ in Eq.~\eqref{eq:corrcoeff} and averages are taken within the bottom layer. 
In thermodynamic equilibrium, the correlation coefficients for these sets of variables should be zero.

\subsubsection{\label{subsubsec:CorrectScheme}Thermodynamically Consistent Case: Using Our Update Scheme}

\begin{figure}
\includegraphics[width=0.8\linewidth]{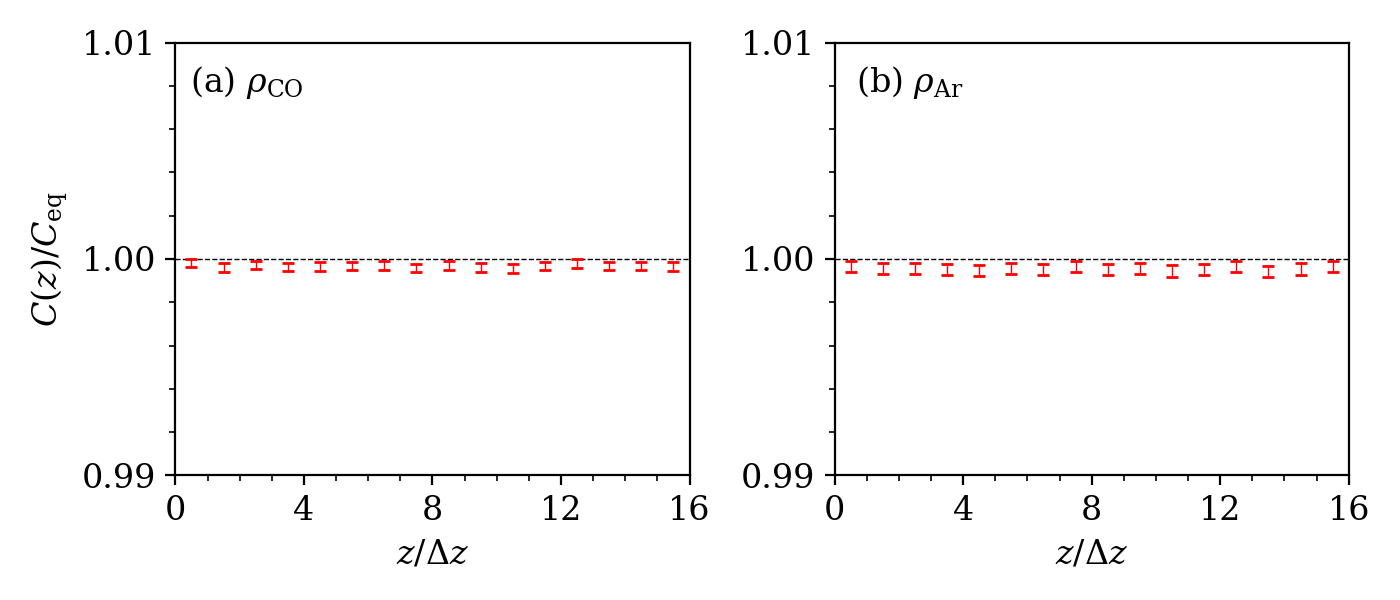}
    \caption{\label{Fig:CorrectCellVariances}
    Cell variances of (a) $\rho_\chem{CO}$ and (b) $\rho_\chem{Ar}$ obtained using our update scheme for $\bar{T}=\SI{800}{K}.$
    The normalized cell variances $C_\phi(z)/C_{\phi,eq}$ are plotted as a function of $z$, where $z = (i-0.5)\Delta z$ is the distance of the $i$th layer ($i=1,\dots,16)$ from the adsorbent surface.
    Error bars show 95\% confidence intervals.}
\end{figure}

Simulation results obtained using our update scheme overall show that it faithfully reproduces the thermodynamic equilibrium.
Figure~\ref{Fig:CorrectCellVariances} shows the cell variance results for $\rho_\mathrm{CO}$ and $\rho_\mathrm{Ar}$ at $\SI{800}{K}$.
For both variables, the profile of $C_\phi(z)$ shows the correct equilibrium value at each layer within 0.1\% error.
The cell variance results for other variables ($v_x$, $v_y$, $v_z$, $T$) also agree with theoretically predicted values within 0.1\% error (see Figure~S1 in the Supplementary Material).
Note that the agreement of our simulation results and theoretically predicted values is remarkable considering that system-size effects which may be present in simulation results due to the finite number of cells in each layer~\cite{KimNonakaBellGarciaDonev2017}, $N_x N_y = 16^2$, are expected to be the order of $1/(N_x N_y) \approx \SI{4e-3}{}$.
For equilibrium simulations at \SI{700}{K}, we observe a similar remarkable agreement for each variable (see Figure~S2 in the Supplementary Material).
 
\begin{figure}
\includegraphics[width=\linewidth]{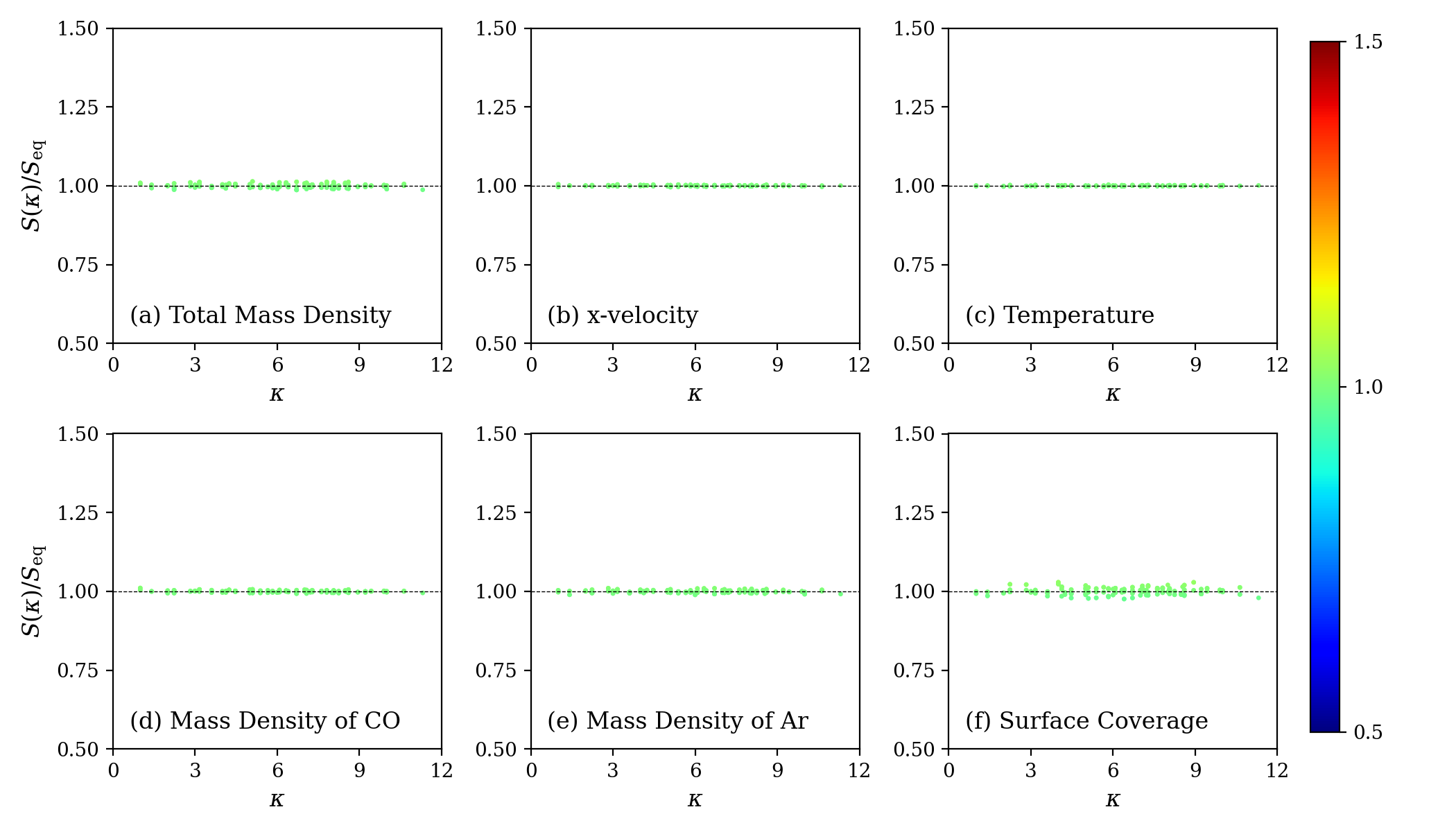}
    \caption{\label{Fig:CorrectStructureFactor}
    Structure factor spectra obtained using our update scheme for $\bar{T}=\SI{800}{K}$.
    The results for the bottom layer contacting with the adsorbent surface are shown: (a) total mass density, (b) $x$-velocity component, (c) temperature, (d) mass density of \chem{CO}, (e) mass density of \chem{Ar}, and (f) surface coverage.
    The normalized structure factors $S_\phi(\kappa)/S_{\phi,eq}$ are plotted as a function of $\kappa=\sqrt{\kappa_x^2+\kappa_y^2}$, where $\kappa_\alpha=k_\alpha(2\pi/L_\alpha)^{-1}$ is the wave index in the $\alpha$-direction ($\alpha=x,y$).}
\end{figure}

The structure factor results also support that our numerical method is thermodynamically consistent.
Figure~\ref{Fig:CorrectStructureFactor} shows the structure factor spectra of $\rho$, $v_x$, $T$, $\rho_\chem{CO}$, $\rho_\chem{Ar}$, and $\theta$ for the bottom layer contacting the adsorbent surface for \SI{800}{K}.
We observe that the spectrum of each variable is flat with the correct value predicted by equilibrium statistical mechanics, showing that there are no unphysical correlations among cells in the bottom layer.
We confirm that the structure factor spectra of the other layers are also flat with the correct values.  
We observe a similar agreement for equilibrium simulations at \SI{700}{K} (see Figure~S3 in the Supplementary Material).
The structure factor spectra of the normal velocity component $v_z$ at $z=\Delta z$ (i.e., $v_z$ on faces between the first and second bottom layers) are also flat with correct values for both \SI{700}{K} and \SI{800}{K} (see Figure~S8 in the Supplementary Material).

The correlation coefficient results also show good agreement with theoretical prediction (i.e., $r_{\phi,\phi'}=0$).
We discuss them in detail in Section~\ref{subsubsec:WithoutEnergyUpdateScheme}, where we compare our simulation results with those obtained using a thermodynamically inconsistent setting, where the energy correction term $\frac12 k_B T$ is not included in the reversible adsorption update.

\subsubsection{\label{subsubsec:MeanScheme}Thermodynamically Inconsistent Case: Using Mean Partial Pressure and Temperature for Adsorption Rate}

We consider here a thermodynamically \emph{inconsistent} setting, where the adsorption rate is evaluated using the mean (or equilibrium) partial pressure of species $\chem{CO}$ and the mean temperature, $\bar{p}_\chem{A}$ and $\bar{T}$, instead of the instantaneous (i.e., fluctuating) partial pressure and temperature as in Eq.~\eqref{eq:lambdaadFHD}.
In other words, the mean rate of adsorption events is replaced with
\begin{equation}
\label{eq:inconsistentrate}
    \lambda_a = k_a(\bar{T})\:\bar{p}_\chem{CO}\:(1-\theta)\: N_{tot}.
\end{equation}
Note that the instantaneous surface coverage $\theta$ is used in Eq.~\eqref{eq:inconsistentrate}.
This setting corresponds to an FHD--KMC coupling, where the KMC solver uses the mean (or equilibrium) values for the hydrodynamic state of the FHD solver to determine the rates of individual KMC events.
Although it may seem reasonable to use the mean values, particularly if fluctuations are relatively small (less than 2\% for $\delta p_\chem{A}$ and less than 1\% for $\delta T$ in our equilibrium simulations), we demonstrate below that this setting causes thermodynamic inconsistency.

\begin{figure}
\includegraphics[width=0.8\linewidth]{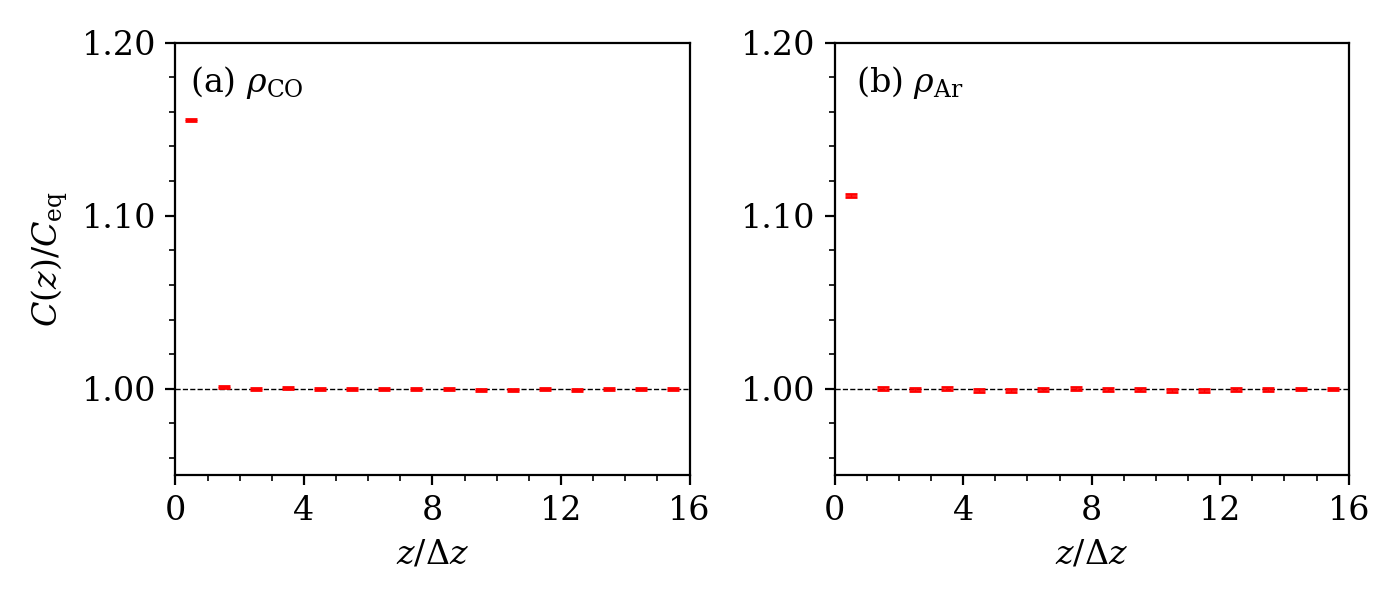}
    \caption{\label{Fig:MeanCellVariances}
    Cell variances of (a) $\rho_\chem{CO}$ and (b) $\rho_\chem{Ar}$ obtained using the thermodynamically inconsistent setting, where the mean partial pressure and temperature are used to evaluate the adsorption rate, see Eq.~\eqref{eq:inconsistentrate}.
    The normalized cell variances $C_\phi(z)/C_{\phi,eq}$ are plotted as a function of $z$, where $z = (i-0.5)\Delta z$ is the distance of the $i$th layer ($i=1,\dots,16)$ from the adsorbent surface.
    Simulation results for $\bar{T}=\SI{800}{K}$ are shown.
    Error bars show 95\% confidence intervals.
    Note that the corresponding plots obtained using our thermodynamically consistent scheme are shown in Figure~\ref{Fig:CorrectCellVariances} with a finer vertical scale.}
\end{figure}

Figure~\ref{Fig:MeanCellVariances} shows the cell variance profiles $C_\phi (z)$ for $\phi = \rho_\chem{CO}$ and $\rho_\chem{Ar}$ at \SI{800}{K}.
Significant deviations (greater than 10\%) from the theoretical values are observed for both variables at the bottom layer contacting the adsorbent surface.
Other variables ($v_x$, $v_y$, $v_z$, $T$) are indirectly affected and small deviations (less that 1\%) are observed at the bottom layer (see Figure~S4 in the Supplementary Material).
These results indicate that thermodynamic inconsistency is caused by the incorrect reversible adsorption update.
Similar observations are made for the cell variance results at \SI{700}{K} (see Figure~S5 in the Supplementary Material).
When the results at \SI{700}{K} and \SI{800}{K} are compared, deviations at the bottom layer are more pronounced at \SI{800}{K}, which is attributed to the larger value of $\bar{r}=\bar{k}_d\bar{\theta}$ at \SI{800}{K}.

\begin{figure}
\includegraphics[width=\linewidth]{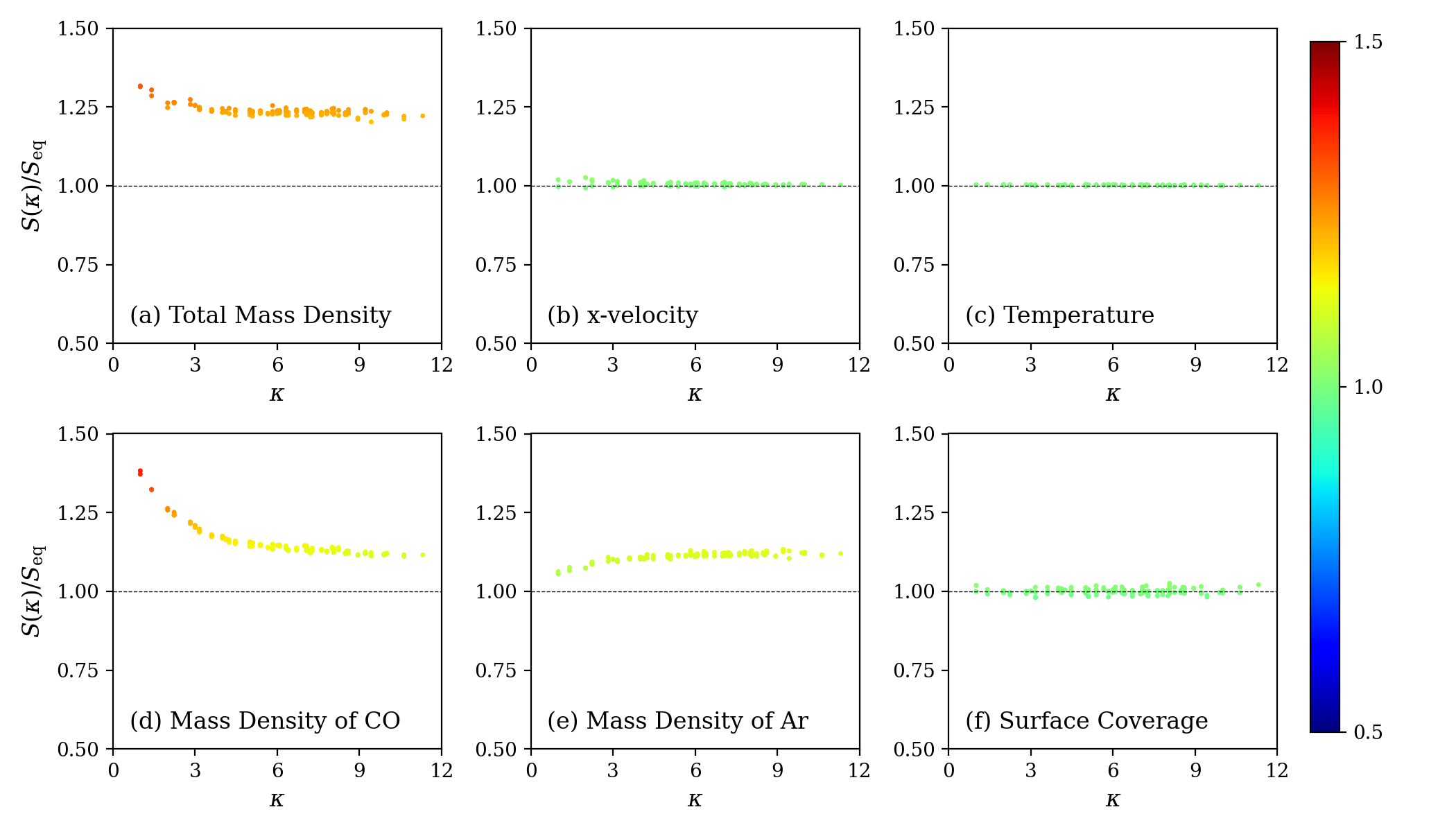}
    \caption{\label{Fig:MeanStructureFactor}
    Structure factor spectra obtained using the thermodynamically inconsistent setting, where the mean partial pressure and temperature are used to evaluate the adsorption rate, see Eq.~\eqref{eq:inconsistentrate}.
    The results for the bottom layer contacting with the adsorbent surface are shown: (a) total mass density, (b) $x$-velocity component, (c) temperature, (d) mass density of \chem{CO}, (e) mass density of \chem{Ar}, and (f) surface coverage.
    The normalized structure factors $S_\phi(\kappa)/S_{\phi,eq}$ are plotted as a function of $\kappa=\sqrt{\kappa_x^2+\kappa_y^2}$, where $\kappa_\alpha=k_\alpha(2\pi/L_\alpha)^{-1}$ is the wave index in the $\alpha$-direction ($\alpha=x,y$).
    Simulation results for $\bar{T}=\SI{800}{K}$ are shown.
    Note that the corresponding plots obtained using our thermodynamically consistent scheme are shown in Figure~\ref{Fig:CorrectStructureFactor}.}
\end{figure}

Figure~\ref{Fig:MeanStructureFactor} shows the structure factor spectra for the bottom layer contacting the adsorbent surface for \SI{800}{K}.
Significant deviations from the theoretical values are observed in the mass density variables, $\rho$, $\rho_\chem{CO}$, and $\rho_\chem{Ar}$. 
Particularly, deviations in the structure factor spectrum of the reactive species $\rho_\chem{CO}$ become larger than 30\% at smaller wave numbers.
Compared with the thermodynamically consistent simulation results shown in Figure~\ref{Fig:CorrectStructureFactor}, changes in the structure factor spectra of the other variables, $v_x$, $T$, and $\theta$ are not noticeable.
The structure factor spectrum of the normal velocity component $v_z$ at $z=\Delta z$ (i.e., $v_z$ on faces between the first and second bottom layers) exhibits minor deviations at smaller waver numbers (see Figure~S8 in the Supplementary Material).
Similar trends are observed for \SI{700}{K} (see Figure~S6 in the Supplementary Material).
As mentioned above, due to the smaller value of $\bar{r}$, deviations caused by the incorrect reversible adsorption update become weaker.
Although rather weak, deviations are noticeable in the second bottom layer (Figure~S7 in the Supplementary Material).
Hence, replacing instantaneous hydrodynamic variables with their mean values leads to thermodynamically inconsistent fluctuation behaviors, especially for the mass density variables near the surface.

\subsubsection{\label{subsubsec:WithoutEnergyUpdateScheme}Thermodynamically Inconsistent Case: When the Energy Correction Term is Not Included}

We finally consider another thermodynamically inconsistent setting, where the energy correction term $\frac12 k_B T$ (see Eq.~\eqref{eq:FHDupdate_c}) is not included in the reversible adsorption update.
By comparing the simulation results obtained using this setting with those obtained using our thermodynamically consistent simulation method, we demonstrate that the energy correction term is needed to reproduce the correct thermodynamic equilibrium.

\begin{figure}
\includegraphics[width=\linewidth]{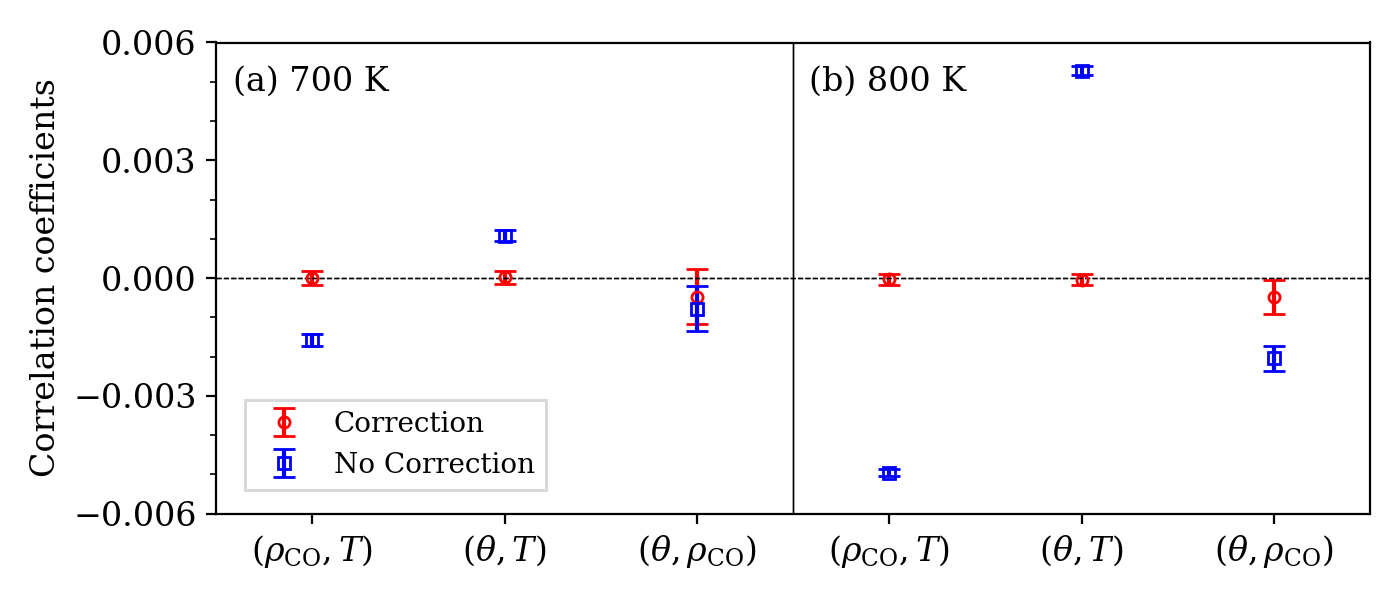}
    \caption{\label{Fig:CorrelationCoefficients}
    The correlation coefficients (see the definition in Eq.~\eqref{eq:corrcoeff}) between the surface coverage ($\theta$), mass density ($\rho_{\mathrm{CO}}$), and temperature ($T$) are shown with error bars depicting 95\% confidence interval for (a) \SI{700}{K} and (b) \SI{800}{K}.
    The red circles (corresponding to `Correction') show the results obtained using our thermodynamically consistent numerical method, whereas the blue squares (`No Correction') show the results obtained using the thermodynamically inconsistent setting, where the energy correction term is not included in the reversible adsorption update.}
\end{figure}

The cell variance and structure factor results (see Figures~S8--S12 in the Supplementary Material) show some noticeable deviations from theoretical prediction.
However, these deviations are not as significant as observed in Section~\ref{subsubsec:MeanScheme}.  
We then compute the correlation coefficients $r_{\phi,\phi'}$ for pairs of thermodynamic variables for the bottom layer, which show how thermodynamic inconsistency develops when the energy correction term is not included.
Figure~\ref{Fig:CorrelationCoefficients} shows the correlation coefficients of $(\rho_\chem{CO},T)$, $(\theta,T)$, and $(\theta,\rho_\chem{CO})$ for \SI{700}{K} and \SI{800}{K}.
Contrary to our numerical method, which gives the correct zero correlation values within statistical errors, the reversible adsorption update without the energy correction term leads to statistically significant nonzero correlations between these thermodynamic variables.
These nonzero correlations are larger at \SI{800}{K} than \SI{700}{K}, which is consistent with the discussion in Section~\ref{subsubsec:MeanScheme}.

\section{\label{sec:Conclusion}Conclusion}

The complexity inherent in modeling reactive gas-solid interfacial systems, particularly at mesoscales where thermal fluctuations are significant and gas hydrodynamics and surface dynamics operate on comparable time and length scales, necessitates a robust, concurrently coupled hybrid simulation approach. 
To address this need, we developed a novel mesoscopic stochastic modeling method that integrates the Langmuir adsorption model with compressible fluctuating hydrodynamics (FHD). 
A primary theoretical achievement of this work was the derivation of a thermodynamically consistent mass--energy update scheme to handle the exchange of mass and energy variables between the gas and surface subsystems during adsorption and desorption events. 
Through a rigorous stochastic analysis applied to the ideal Langmuir model and the full hydrodynamic system, we analytically confirmed that this update scheme successfully captures the thermodynamic equilibrium predicted by equilibrium statistical mechanics.

A crucial element identified during the derivation of the mass–energy update scheme was the requirement for an internal energy correction term, quantified as $\frac12 k_B T$ per molecule. 
This correction is necessary because the mean kinetic energy of a gas molecule colliding with the surface differs from the mean kinetic energy calculated from the bulk Maxwell--Boltzmann distribution.  
Specifically, the normal velocity component of a molecule colliding with the wall adheres to a Rayleigh distribution, resulting in an average kinetic energy of $k_B T$ for that component, rather than $\frac12 k_B T$ associated with the other two velocity components; hence, the mean kinetic energy of the molecule ($2 k_B T$) is larger by $\frac12 k_B T$ than the standard $\frac32 k_B T$ mean kinetic energy for the bulk gas.
Furthermore, we developed a thermodynamically consistent reaction (TCR) model for Langmuir adsorption, which guaranties that the formulation and parameter selection are based on consistent chemical potential models, thus ensuring that the relationship between the equilibrium constant and the rate constants is preserved.

We performed extensive numerical validations using equilibrium simulations of an ideal gas mixture (\chem{CO} and \chem{Ar}, with \chem{CO} undergoing reversible adsorption) to confirm the accuracy of our methodology beyond the weak-noise limit. 
The simulation  results obtained using our update scheme faithfully reproduced the expected thermodynamic equilibrium properties.
Specifically, cell variances and structure factor spectra for all state variables, including mass densities, velocity components, temperature, and surface coverage, agreed with theoretical predictions based on equilibrium statistical mechanics within minimal statistical errors, e.g., cell variance profiles agreeing within 0.1\% error. 
Conversely, thermodynamically inconsistent settings---such as replacing instantaneous partial pressure and temperature with mean values in the adsorption rate calculation (mimicking a passive macro-micro coupling)---led to significant deviations, e.g., exceeding 10\% error in cell variances and unphysical fluctuations for mass density variables near the adsorbent surface.
Most importantly, our validation study demonstrated the critical role of the $\frac12 k_B T$ energy correction term. 
When this correction was omitted from the reversible adsorption update, the resulting simulation displayed statistically significant nonzero correlations between thermodynamic variables such as $(\rho_\chem{CO}, T)$, $(\theta, T)$, and $(\theta, \rho_\chem{CO})$ in the bottom layer contacting the adsorbent surface, which should be zero at equilibrium.
These findings confirm that our methodology provides a foundational, thermodynamically consistent framework for modeling fluctuations at the gas-solid interface. 

As mentioned in the Introduction, our mass--energy update scheme is designed as a direct precursor for the development of a promising two-way, concurrent hybrid approach, namely FHD--KMC coupling, for reactive gas-solid interfacial systems at the mesoscale.
Hence, future work includes the algorithmic development and implementation of FHD--KMC coupling.
In addition, extending our mass--energy update scheme to a system with a nonzero mean flow velocity (e.g., flow reactor as opposed to batch reactor) would be an interesting future direction.
Since the collision rate and thus the adsorption rate depend on the flow velocity, we expect that momentum would also need to be included in a thermodynamically consistent reversible adsorption update.

\section*{Supplementary Material}
The supplementary material encompasses the following: simulation results of cell variance profiles and structure factor spectra for thermodynamically consistent case using our update scheme; thermodynamically inconsistent case using mean partial pressure and temperature for adsorption rate; thermodynamically inconsistent case when the energy correction term is not included.

\begin{acknowledgments}
The authors would like to express their gratitude and respect to Dr.~Aleksandar Donev, who left a legacy in the field of theoretical and computational fluctuating hydrodynamics.
This work was supported in part by the National Science Foundation under Grant No.\ CHE-2213368. 
This work was supported in part by the U.S. Department of Energy, Office of Science, Office of Advanced Scientific Computing Research, Applied Mathematics Program under contract No.\ DE-AC02-05CH11231.
H.T.J.\ and H.K.\ acknowledge the support by the National Research Foundation of Korea funded by the Korean government (Nos.\ RS-2024-00405261 and RS-2024-00450102).
This research used resources of the National Energy Research Scientific Computing Center, which is supported by the Office of Science of the U.S.\ Department of Energy under Contract No.\ DE-AC02-05CH11231.
\end{acknowledgments}

\section*{Author Declarations}
{\bf Conflict of Interest}\\
The authors have no conflicts to disclose.

\section*{Data Availability}
The data that support the findings of this study are available within the article and its supplementary material.

\appendix

\section{\label{sec:appendixTCR}Derivation of the Modified Arrhenius Form for $K(T)$}

To derive Eq.~\eqref{eq:TCRKKst}, which shows the temperature dependence of the equilibrium constant $K(T)$ in the form of the modified Arrhenius equation, we start with the chemical potential of each subsystem.
We use dimensionless chemical potentials (per particle), which are normalized by $k_B T$.
The chemical potential of a gas molecule in an ideal gas is given by
\begin{equation}
    \hat{\mu}_g(p,T) = \hat{\mu}_g^\circ(T) + \log\frac{p}{p^{st}},
\end{equation}
where $\hat{\mu}_{g}^\circ(T)$ is the chemical potential at $p = p^{st}$.
Note that $p$ refers to the partial pressure of species \chem{A}.
The chemical potential of an adsorbate on the ideal adsorbent is given by~\cite{Hill1987, ConwayAngersteinKozlowskaDhar1974}
\begin{equation}
    \hat{\mu}_{ads}(\theta,T) = \hat{\mu}_{ads}^\circ(T) + \log\frac{\theta}{1-\theta},
\end{equation}
where $\hat{\mu}_{ads}^\circ$ is the chemical potential at $\theta = \frac12$.
Equating $\hat{\mu}_{ads}$ and $\hat{\mu}_g$, Eq.~\eqref{eq:EquilibConstantDefn}) gives
\begin{equation}
    K(T) = \frac{1}{p^{st}}\exp\left(\hat{\mu}_g^\circ(T)-\hat{\mu}_{ads}^\circ(T)\right).
\end{equation}
By considering the ratio of $K(T)$ to $K(T^{st})$, we express the temperature dependence of $K$ in terms of the chemical potential differences at $T$ and $T^{st}$:
\begin{equation}
\label{eq:KTKTstexpdeltamu}
    K(T) = K(T^{st})
    \frac{\exp\left(\hat{\mu}_g^\circ(T)-\hat{\mu}_g^\circ(T^{st})\right)}{\exp\left(\hat{\mu}_{ads}^\circ(T)-\hat{\mu}_{ads}^\circ(T^{st})\right)}.
\end{equation}

As shown in Ref.~\citenum{PolimenoKimBlanchetteSrivastavaGarciaNonakaBell2025}, one can further reduce the term $\hat{\mu}_g^\circ(T)-\hat{\mu}_g^\circ(T^{st})$ by assuming that the specific heat capacity of the gas at constant pressure, $c_{p,g}$, is constant.
The specific enthalpy and entropy of the gas are given as
\begin{equation}
    h_g(T) = h_g^{st}+c_{p,g} (T-T^{st}),\quad s_g(T) = s_g^{st} + c_{p,g} \log\frac{T}{T^{st}},
\end{equation}
respectively, where $h_g^{st}$ and $s_g^{st}$ are the corresponding values at $T=T^{st}$.
Since the dimensionless chemical potential is given by
\begin{equation}
    \hat{\mu}_g^\circ (T) = \frac{m}{k_B T}\Bigl(h_g(T)-T s_g(T)\Bigr),
\end{equation}
we have
\begin{equation}
\label{eq:mugform}
    \hat{\mu}_g^\circ(T)
    = \frac{m\varepsilon_g}{k_B T} -\frac{m c_{p,g}}{k_B}\log T + \mbox{(temp. indep. terms)},
\end{equation}
where $\varepsilon_g = h_g^{st}-c_{p,g} T^{st}$, and thus,
\begin{equation}
\label{eq:expdeltamug}
    \exp\Bigl(\hat{\mu}_g^\circ(T)-\hat{\mu}_g^\circ(T^{st})\Bigr)
    = \exp\left[\frac{m\varepsilon_g}{k_B} \left(\frac{1}{T}-\frac{1}{T^{st}}\right)\right]\left(\frac{T}{T^{st}}\right)^{-m c_{p,g}/k_B}.
\end{equation}
Note that the specific internal energy of an ideal gas is given as in Eq.~\eqref{eq:eg} because
\begin{equation}
    e_g(T) = h_g(T) - \frac{k_B T}{m}.
\end{equation}

For the ideal adsorbent, we assume that its specific internal energy is given as in Eq.~\eqref{eq:eads}.
By ignoring thermal expansion of the adsorbent, we obtain the specific enthalpy and entropy:
\begin{equation}
    h_{ads}(T) = \varepsilon_{ads} + c_{ads}T, \quad
    s_{ads}(T) = s_{ads}^{st} + c_{ads}\log\frac{T}{T^{st}},
\end{equation}
where $s_{ads}^{st}$ is the specific entropy at $T=T^{st}$.
Following the same procedure as for the gas phase, we express $\hat{\mu}_{ads}^\circ$ as
\begin{equation}
\label{eq:muadsform}
    \hat{\mu}_{ads}^\circ (T) = \frac{m \varepsilon_{ads}}{k_B T} - \frac{m c_{ads}}{k_B} \log T + \mbox{(temp. indep. terms)},
\end{equation}
and obtain
\begin{equation}
\label{eq:expdeltamuads}
    \exp\Bigl(\hat{\mu}_{ads}^\circ(T)-\hat{\mu}_{ads}^\circ(T^{st})\Bigr)
    = \exp\left[\frac{m\varepsilon_{ads}}{k_B} \left(\frac{1}{T}-\frac{1}{T^{st}}\right)\right]\left(\frac{T}{T^{st}}\right)^{-m c_{ads}/k_B}.
\end{equation}
We note that several expressions for the chemical potential of an ideal adsorbent have been derived using different assumptions~\cite{Hill1987, ErpTrinhKjelstrupGlavatskiy2014, KnopfAmmann2021} and they agree with Eq.~\eqref{eq:muadsform}.
By substituting Eqs.~\eqref{eq:expdeltamug} and \eqref{eq:expdeltamuads} into Eq.~\eqref{eq:KTKTstexpdeltamu}, we finally obtain Eq.~\eqref{eq:TCRKKst}.

\section{\label{sec:appendixFHDreactgasmix}FHD Description of a Reactive Gas Mixture}

For an ideal gas mixture of species $\chem{A}$ and $\chem{B}$, we denote the species mass densities by $\rho_\chem{A}$ and $\rho_\chem{B}$, the total mass density by $\rho \equiv \rho_\chem{A}+\rho_\chem{B}$, the fluid velocity by $\boldsymbol{v}$, and the total specific energy (i.e., energy per mass) by $E$.
The time evolution of the species mass densities ($\rho_\chem{A}$ and $\rho_\chem{B}$), momentum density ($\rho\boldsymbol{v}$), and energy density ($\rho E$) is described by the fluctuating Navier--Stokes equations~\cite{PolimenoKimBlanchetteSrivastavaGarciaNonakaBell2025}:
\begin{subequations}
\label{eq_FNS}
\begin{align}
    \label{eq_FNS_rhoA}
    \frac{\partial\rho_\chem{A}}{\partial t} &=  
    -\nabla\cdot(\rho_\chem{A}\boldsymbol{v}) 
    -\nabla\cdot\mathbfcal{F}_\chem{A}
    + m_\chem{A}\omega_\chem{A}, \\
    \label{eq_FNS_rhoB}
    \frac{\partial\rho_\chem{B}}{\partial t} &=  
    -\nabla\cdot(\rho_\chem{B}\boldsymbol{v}) 
    -\nabla\cdot\mathbfcal{F}_\chem{B} 
    + m_\chem{B}\omega_\chem{B}, \\
    \label{eq_FNS_rhou}
    \frac{\partial(\rho\mathbf{u})}{\partial t} &= 
    -\nabla\cdot(\rho\boldsymbol{v}\boldsymbol{v}^T)
    -\nabla{p} 
    - \nabla\cdot\boldsymbol{\Pi}, \\
    \label{eq_FNS_rhoE}
    \frac{\partial(\rho{E})}{\partial t} &= 
    -\nabla\cdot(\rho E\boldsymbol{v} + p\boldsymbol{v}) 
    -\nabla\cdot\boldsymbol{\Phi}
    -\nabla\cdot(\boldsymbol{\Pi}\cdot\boldsymbol{v}).
\end{align}
\end{subequations}
Here, $\mathbfcal{F}_\chem{A}$ and $\mathbfcal{F}_\chem{B}$ are the species mass fluxes and $\omega_\chem{A}$ and $\omega_\chem{B}$ are the production rate of species $\chem{A}$ and $\chem{B}$ due to chemical reactions.
The pressure is denoted by $p$ and $\boldsymbol{\Pi}$, and $\boldsymbol{\Phi}$ are the momentum and heat fluxes, respectively.

While Eqs.~\eqref{eq_FNS} may superficially resemble as the deterministic Navier--Stokes equations, it is important to note that the standard deterministic fluxes for species mass, momentum, and heat are augmented with stochastic components that represent fluctuations.
In other words, these fluxes are expressed as
\begin{equation}
    \mathbfcal{F}_\chem{A} 
    = \overline{\mathbfcal{F}}_\chem{A}  + \widetilde{\mathbfcal{F}}_\chem{A},\quad
    \mathbfcal{F}_\chem{B} 
    = \overline{\mathbfcal{F}}_\chem{B}  + \widetilde{\mathbfcal{F}}_\chem{B},\quad
    \boldsymbol{\Pi} = \overline{\boldsymbol{\Pi}} + \widetilde{\boldsymbol{\Pi}},\quad
    \boldsymbol{\Phi} = \overline{\boldsymbol{\Phi}} + \widetilde{\boldsymbol{\Phi}},
\end{equation}
where the overline and tilde notations denote the deterministic and stochastic parts, respectively.
For the explicit forms of these fluxes, we refer the reader to Refs.~\onlinecite{BalakrishnanGarciaDonevBell2014, SrivastavaLadigesNonakaGarciaBell2023}.
In a similar fashion, the chemical production rates are also expressed as the sum of deterministic and stochastic parts:
\begin{equation}
    \omega_\chem{A} = \overline{\omega}_\chem{A} + \widetilde{\omega}_\chem{A}, \quad
    \omega_\chem{B} = \overline{\omega}_\chem{B} + \widetilde{\omega}_\chem{B}
\end{equation}
For the explicit forms of the chemical production rates, we refer the reader to Refs.~\onlinecite{PolimenoKimBlanchetteSrivastavaGarciaNonakaBell2025}.

The relation between the total specific energy $E$ and the temperature $T$ is given by
\begin{equation}
\label{eq:Edef}
    E = \frac{1}{2}|\boldsymbol{v}|^2 + e(T,\rho_\chem{A},\rho_\chem{B}).
\end{equation}
Here, the total specific internal energy $e(T,\rho_\chem{A},\rho_\chem{B})$ is a function of temperature and chemical composition.
For an ideal gas mixture, one can simply express $e$ as the weighted sum of the specific internal energy of each species:
\begin{equation}
\label{eq:edef}
    e(T,\rho_\chem{A},\rho_\chem{B}) =
    \frac{1}{\rho} \Bigl\{ \rho_\chem{A} e_\chem{A}(T) + \rho_\chem{B} e_\chem{B}(T) \Bigr\}.
\end{equation}
From the constant specific heat capacity assumption, we set 
\begin{equation}
\label{eq:eAeBdef}
    e_\chem{A}(T) = \varepsilon_\chem{A} + c_{v,\chem{A}} T,\quad
    e_\chem{B}(T) = \varepsilon_\chem{A} + c_{v,\chem{B}} T,
\end{equation}
and thus Eqs.~\eqref{eq:Edef}--\eqref{eq:eAeBdef} give 
\begin{equation}
\label{eq:Eexpression}
    E(\rho_\chem{A},\rho_\chem{B},\boldsymbol{v},T) = \frac12 |\boldsymbol{v}|^2 
    + \frac{\rho_\chem{A}\varepsilon_\chem{A}+\rho_\chem{B}\varepsilon_\chem{B}}{\rho}
    + \frac{\rho_\chem{A}c_{v,\chem{A}}+\rho_\chem{B}c_{v,\chem{B}}{\rho}}{\rho} T.
\end{equation}


%

\end{document}